\documentclass{aastex631}
\usepackage{CJK}
\usepackage[whole]{bxcjkjatype}
\usepackage{comment}

\received{2024 November 12}
\revised{2025 January 22}
\accepted{2025 January 24}

\shorttitle{Supersonic cloud-cloud collisions in RCW 106}
\shortauthors{Kohno et al.}
\graphicspath{{./}{figures/}}
\def\r{\bibitem[]{}} 
 \def\be{\begin{equation}} 
 \def\ee{\end{equation}}
\def\kms{~km~s$^{-1}$~}

\begin{document}
\begin{CJK*}{UTF8}{gbsn}
\title{Giant Molecular clouds in RCW 106 (G333): Galactic mini-starbursts and massive star formation induced by supersonic cloud-cloud collisions }

\author[0000-0003-1487-5417]{Mikito Kohno (河野樹人)}
\affiliation{Curatorial Division, Nagoya City Science Museum, 2-17-1 Sakae, Naka-ku, Nagoya, Aichi 460-0008, Japan}
\affiliation{Department of Physics, Graduate School of Science, Nagoya University, Furo-cho, Chikusa-ku, Nagoya, Aichi 464-8602, Japan}

\author[0000-0002-1865-4729]{Rin I. Yamada (山田麟)}
\affiliation{Department of Physics, Graduate School of Science, Nagoya University, Furo-cho, Chikusa-ku, Nagoya, Aichi 464-8602, Japan}

\author[0000-0002-1411-5410]{Kengo Tachihara (立原研悟)}
\affiliation{Department of Physics, Graduate School of Science, Nagoya University, Furo-cho, Chikusa-ku, Nagoya, Aichi 464-8602, Japan}

\author[0000-0002-6375-7065]{Shinji Fujita (藤田真司)}
\affiliation{Institute of Statistical Mathematics, 10-3 Midori-cho, Tachikawa, Tokyo, Japan}

\author[0000-0003-2735-3239]{Rei Enokiya (榎谷玲依)}
\affiliation{Faculty of Science and Engineering, Kyushu Sangyo University, 2-3-1 Matsukadai, Fukuoka 813-8503, Japan}

\author[0000-0002-2062-1600]{Kazuki Tokuda (徳田一起)}
\affiliation{Department of Earth and Planetary Sciences, Faculty of Sciences, Kyushu University, Fukuoka, Fukuoka 819-0395, Japan}
\affiliation{National Astronomical Observatory of Japan (NAOJ), National Institutes of Natural Sciences (NINS), 2-21-1 Osawa, Mitaka, Tokyo 181-8588, Japan}

\author{Asao Habe (羽部朝男)}
\affiliation{Department of Physics, Faculty of Science, Hokkaido University, N10 W8, Kitaku, Sapporo, Hokkaido 060-0810, Japan}

\author[0000-0003-2062-5692]{Hidetoshi Sano (佐野栄俊)}
\affiliation{Faculty of Engineering, Gifu University, 1-1 Yanagido, Gifu, Gifu 501-1193, Japan}
\affiliation{Center for Space Research and Utilization Promotion (c-SRUP), Gifu University, 1-1 Yanagido, Gifu 501-1193, Japan}

\author[0000-0003-0324-1689]{Takahiro Hayakawa (早川貴敬)}
\affiliation{Department of Physics, Graduate School of Science, Nagoya University, Furo-cho, Chikusa-ku, Nagoya, Aichi 464-8602, Japan}

\author[0009-0002-0025-1646]{Fumika Demachi (出町史夏)}
\affiliation{Department of Physics, Graduate School of Science, Nagoya University, Furo-cho, Chikusa-ku, Nagoya, Aichi 464-8602, Japan}

\author{Takuto Ito (伊藤拓冬)}
\affiliation{Department of Physics, Graduate School of Science, Nagoya University, Furo-cho, Chikusa-ku, Nagoya, Aichi 464-8602, Japan}

\author[0000-0002-2794-4840]{Kisetsu Tsuge (柘植紀節)}
\affiliation{Faculty of Engineering, Gifu University, 1-1 Yanagido, Gifu, Gifu 501-1193, Japan}
\affiliation{Institute for Advanced Study, Gifu University, 1-1 Yanagido, Gifu, Gifu 501-1193, Japan}
\affiliation{Institute for Advanced Research, Nagoya University, Furo-cho, Chikusa-ku, Nagoya, Aichi 464-8601, Japan}

\author[0000-0003-0732-2937]{Atsushi Nishimura (西村淳)}
\affiliation{Nobeyama Radio Observatory, National Astronomical Observatory of Japan (NAOJ), National Institutes of Natural Sciences (NINS), 462-2, Nobeyama, Minamimaki, Minamisaku, Nagano 384-1305, Japan}

\author[0000-0003-3990-1204]{Masato I.N. Kobayashi (小林将人)}
\affiliation{I. Physikalisches Institut Universit\UTF{00E4}t zu K\UTF{00F6}ln Z\UTF{00FC}lpicher Stra\UTF{00DF}e 77 50937 K\UTF{00F6}ln Germany}

\author[0000-0001-5792-3074]{Hiroaki Yamamoto (山本宏昭)}
\affiliation{Department of Physics, Graduate School of Science, Nagoya University, Furo-cho, Chikusa-ku, Nagoya, Aichi 464-8602, Japan}

\author[0000-0002-8966-9856]{Yasuo Fukui (福井康雄)}
\affiliation{Department of Physics, Graduate School of Science, Nagoya University, Furo-cho, Chikusa-ku, Nagoya, Aichi 464-8602, Japan}

\correspondingauthor{Mikito Kohno}
\email{kohno.ncsmp@gmail.com,mikito.kohno@gmail.com}


\begin{abstract}
{To reveal} the origin of the mini-starbursts in the Milky Way,
we carried out large-scale CO observations toward the RCW 106 giant molecular cloud (GMC) complex using the NANTEN2 4-m radio telescope operated by Nagoya University.
{We also analyzed the Mopra Southern Galactic plane CO survey and Herschel infrared continuum archival data.}
{The RCW 106 GMC complex contains the radial velocity components of $-68$\kms and $-50$\kms reported by \citet{2015ApJ...812....7N}. 
Focusing on the RCW 106 East and West region with the massive star formation having the bright infrared dust emission, we found that these regions have three different velocity components with $\sim$ 10\kms differences.}
{{The two out of three} velocity components show morphological correspondence with the infrared cold dust emission and connect with the bridge feature on a position-velocity diagram.}
Therefore, {two} molecular clouds {(MCs)} with $\sim$ 10\kms differences are likely to be physically associated with massive star-forming regions in the GMC complex.
Based on these observational results, {we argue that mini-starbursts and massive star/cluster formation in the RCW 106 GMC complex are induced by supersonic cloud-cloud collisions in an agglomerate of molecular gas on the Scutum-Centaurus arm.}
\end{abstract}

\keywords{Interstellar medium (847) --- Giant molecular clouds (653) ---  Interstellar clouds (834) --- Milky Way disk (1050) --- Molecular gas (1073) --- CO line emission (262)}

\section{Introduction}
\begin{figure*}[h]
\begin{center} 
\includegraphics[width=19cm]{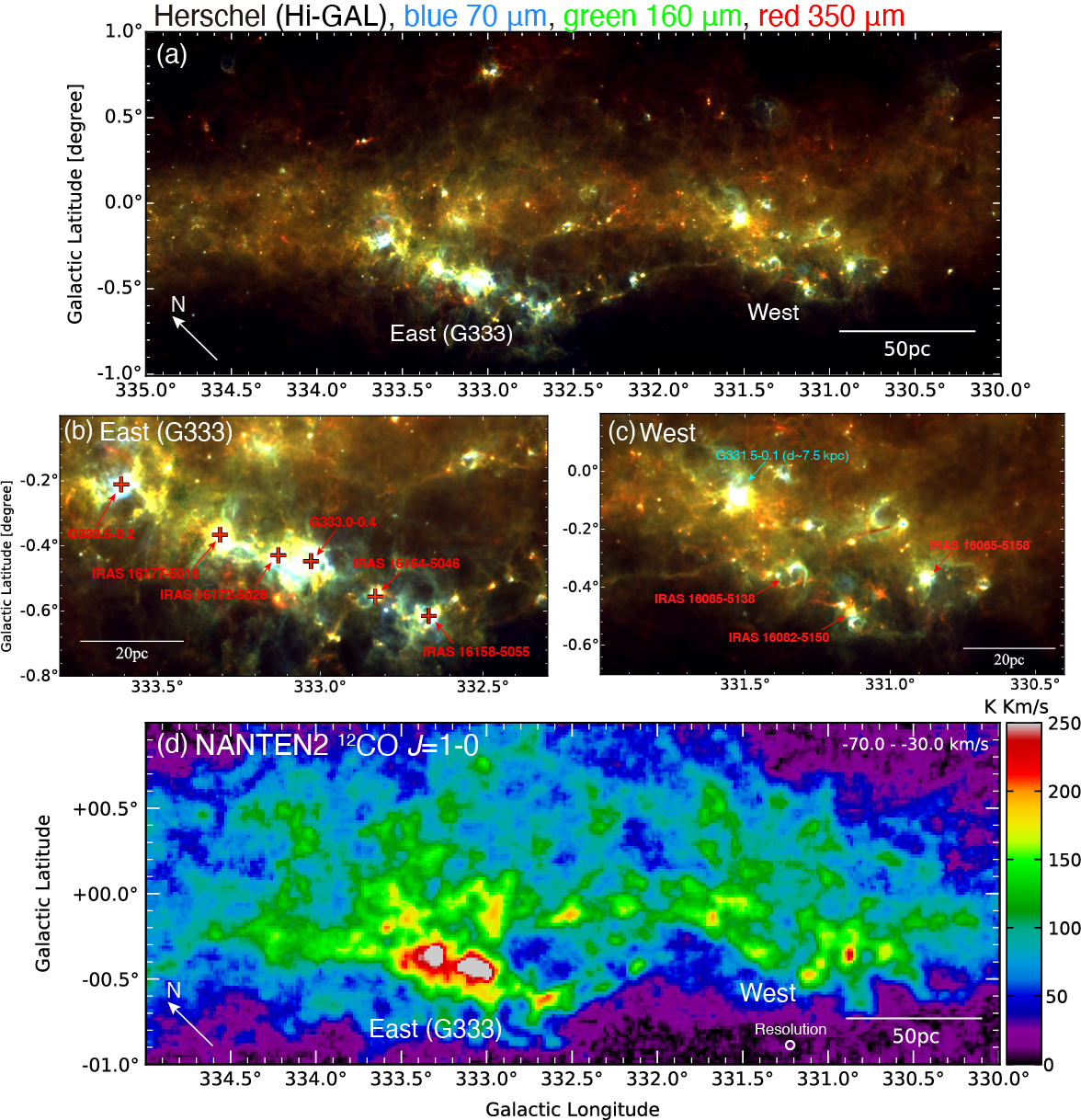}
\end{center} 
\caption{(a) The Herschel three-color composite image of the RCW 106 GMC complex. Blue, green, and red show the Herschel/PACS 70 $\mu$m, Herschel/PACS 160 $\mu$m, and Herschel/SPIRE 350 $\mu$m distributions, respectively \citep{2016A&A...591A.149M}. {(b) The Herschel close-up image of RCW 106 East (G333). The red cross marks are the positions of the massive star-forming regions in RCW 106 East \citep{2006MNRAS.367.1609B}. (c) The Herschel close-up image of RCW 106 West.} (d) The NANTEN2 $^{12}$CO~$J$~=~1--0 integrated intensity map of the RCW 106 GMC complex. The integrated velocity range is from $-70$\kms to $-30$\kms. }
\label{colorimg}
\end{figure*}
Massive star formation is an essential topic in astrophysics because massive stars emit ultraviolet radiation and affect the surrounding interstellar medium. 
They are born as star clusters in giant molecular clouds (GMCs) on the Galactic disk (e.g., \citealp{2003ARA&A..41...57L,2018ASSL..424....1A,2019ARA&A..57..227K,2020SSRv..216...64K}).
Extreme massive star formation in a GMC is known as mini-starburst regions in the Milky Way, which have been studied by {multi-wavelength observations, for example, radio,  infrared, {optical (H${\alpha}$)}, X-ray, and gamma ray} (\citealp{2011A&A...529A..41N}; {Russeil et al. 2016}; \citealp{2018PASJ...70S..41F,2018ARA&A..56...41M,2020A&A...640A..60Y,2021PASJ...73S.129K,2023A&A...674A..76P,2024A&A...682A.172S,2024arXiv240809905L}).

RCW 106 \citep{1960MNRAS.121..103R} is the massive star-forming region associated with the Scutum-Centaurus arm at a kinematic distance of $\sim 3.6$ kpc from the solar system \citep{2015ApJ...812....7N}. This kinematic distance is calculated by the radial velocity of molecular gas using the Galactic rotation curve \citep{2009ApJ...700..137R} and is also consistent with the spectrophotometric distance at the OB star cluster in RCW 106 \citep{2011MNRAS.411..705M}.
The total molecular mass is $5.9 \times 10^6$ $M_{\odot}$ as one of the {most massive} molecular cloud complexes in the Milky Way \citep{2015ApJ...812....7N,2016ApJ...833...23N}.
The RCW 106 GMC complex contains $\sim 50$ O-type stars \citep{2015ApJ...812....7N} and has been studied as the mini-starburst region by infrared, {optical (H${\alpha}$),} and radio observations (e.g., \citealp{1977A&A....60..221G,2001MNRAS.326..293K,2004A&A...426..119M,2005A&A...429..497R,2006MNRAS.367.1609B,2007MNRAS.377..491B,2008MNRAS.386.1069W,2009MNRAS.395.1021L,2011MNRAS.415..525L,2012MNRAS.419..211S,2014MNRAS.441..256L,2016MNRAS.458.3429W,2019ApJ...875L..16T,2022A&A...662A...8M}).
Previous CO observations revealed that molecular clouds are composed of different radial velocities of $-90$\kms, $-64$\kms, and $-48$\kms \citep{2015ApJ...812....7N}.
The $-90$\kms component is associated with the massive star-forming region G331.5-0.1 (MCC 331-90) at a distance of $\sim 7.5$ kpc in the Norma arm \citep{2013ApJ...774...38M}. 
\cite{2015ApJ...812....7N} suggest that the $-64$\kms and $-48$\kms components are associated with the RCW 106 GMC complex in the Scutum-Centaurus arm.

Figure \ref{colorimg}(a) presents a three-color composite image obtained by the Herschel infrared Galactic Plane Survey (Hi-GAL). Blue, green, and red presents $70\ \mu$m, $160\ \mu$m, and $350\ \mu$m continuum image, respectively. These far-infrared image mainly trace cold dust components with the temperature of $10<T<40$ K (e.g., \citealp{2010PASP..122..314M}).

RCW 106 complex has two bright regions in the infrared image, as shown in the close-up image in Figures \ref{colorimg}(b) and \ref{colorimg}(c); hereafter, we call RCW 106 East and RCW 106 West, respectively. 
RCW 106 East (G333) is the brightest region in the GMC complex and contains the massive cluster G333.6--0.2 \citep{1973ApJ...182L.125B,1980ApJ...241..709H,1998MNRAS.296..225F,2001MNRAS.327..233F,2005MNRAS.356..801F,2006MNRAS.368.1843F,2013A&A...558A.119K,2014A&A...563A.123G,2024ApJ...960...48T,2024A&A...687A.163B,2024A&A...687A.217D}, massive star-forming region IRAS 16177--5018, IRAS 16172--5028 {(=G333.1--0.4;\citealp{2005AJ....129.1523F})}, G333.0--0.4 \citep{2015MNRAS.453.3245L}, IRAS 16164--5046 {(=G332.8--0.6)}, and IRAS 16158--5055 (see Figure 3 in \citealp{2006MNRAS.367.1609B}).
RCW 106 West includes massive star-forming regions of IRAS 16085--5138 \citep{2012MNRAS.423.2425P}, IRAS 16082--5150, and IRAS 16065--5158 \citep{2011A&A...526A..59D}.
The region between RCW 106 East and RCW 106 West is the G332 region reported the filamentary structure \citep{2019MNRAS.484.2089R}.
Recently, \cite{2023A&A...676A..69Z} performed large-scale $^{13}$CO~($J$~=~3--2) observations toward the RCW 106 GMC complex using the APEX/LAsMA heterodyne camera. They reported the hub-filament system in RCW 106 and suggested the state of global gravitational collapse (see also \citealp{2024A&A...682A.128Z,2024arXiv240313442Z}). 
On the other hand, previous studies have not clarified the relation between two radial velocity components in the RCW 106 GMC complex and massive star/cluster formation.
{It has also been suggested that shock waves resulting from large-scale gas collisions are crucial in the formation of hub-filament systems (e.g., \citealp{2018PASJ...70S..53I,2024arXiv240806826M,2024arXiv241113870M}), and observationally, massive star-forming regions consistent with these predictions are being discovered (e.g., \citealp{2019ApJ...886...14F,2019ApJ...886...15T,2022ApJ...933...20T,2023ApJ...955...52T}). Therefore, investigating the global gas kinematics in regions with hub-filamentary cloud candidates is becoming an urgent issue. RCW 106, which is one of the most massive complexes in the Galactic plane observable from the southern sky, is thus an optimal region for such research.}
In this paper, {we aim to reveal the relation between the origin of massive star formation and different radial velocity components in the RCW 106 GMC complex}. 

This paper is structured as follows: section 2 introduces observations and the archival data; Section 3 shows the results; Section 4 discusses the massive cluster formation scenario in the RCW 106 GMC complex; and Section 5 summarizes this paper.


\section{Observations and archival data}

\begin{table*}[h]
\caption{Observational data properties.}
\begin{tabular}{cccccccccc}
\hline
\multicolumn{1}{c}{Telescope} & Line  & HPBW & Grid &  Velocity & r.m.s noise$^*$ & References \\
& & &  &Resolution & level &\\
\hline
NANTEN &$^{12}$CO~$J$~=~1--0  & 
{$2\farcm6$} & 4\arcmin & 0.65 km s$^{-1}$ & $\sim 0.2$ K  & [1,2]\\
NANTEN2 &$^{12}$CO~$J$~=~1--0 & {$2\farcm6$} & 1\arcmin & 0.16 km s$^{-1}$& $\sim 1.1$ K  &  \\
Mopra &$^{12}$CO~$J$~=~1--0 & $\sim 33$\arcsec  & 30\arcsec &0.1 km s$^{-1}$& $\sim 0.8 $ K & [3,4,5,6] \\
 &$^{13}$CO~$J$~=~1--0 &  $\sim 33$\arcsec  &30\arcsec & 0.1 km s$^{-1}$ & $\sim 0.3$ K  & [3,4,5,6] \\
\hline
\hline
Telescope/Survey & Band  & Detector &Resolution& Grid & & References  & \\
\hline
{Herschel}/Hi-GAL & 70 $\mu$m & PACS  &$\sim$ 6\arcsec & 3\arcsec & & [7,8] & \\
{Herschel}/Hi-GAL  & 160 $\mu$m & PACS  & $\sim$ 12\arcsec & 4.5\arcsec  & & [7,8]  &\\
{Herschel}/Hi-GAL  & 350 $\mu$m  & SPIRE  &$\sim$ 24\arcsec & 8\arcsec & &  [7,9]  &\\
\hline
\end{tabular}
\label{obs_param}
 \vspace{3pt}
\label{obs_para}\\
{\raggedright Notes. $^*$The r.m.s noise level is taken from the final cube data using this paper. References [1] \citet{2004ASPC..317...59M},[2] \citet{2010PASJ...62..557T},[3] \citet{2013PASA...30...44B}, [4] \citet{2015PASA...32...20B} [5] \citet{2018PASA...35...29B} [6] \citet{2023PASA...40...47C}, [7] \citet{2010PASP..122..314M}[8] \citet{2010A&A...518L...2P}, [9] \citet{2010A&A...518L...3G}
 \par}
\end{table*}

\subsection{NANTEN and NANTEN2 $^{12}$CO~$J$~=~1--0 observations}
We carried out the $^{12}$CO~$J$~=~1--0 survey observations toward the southern Galactic plane from March 2012 to December 2012 using the NANTEN2 radio telescope at Pampalabola in Chile. 
NANTEN2\footnote{\url{https://www.a.phys.nagoya-u.ac.jp/jp/index.php/nanten2/}} has a diameter of 4 m, operated by Nagoya University.
The half power beam width (HPBW) is $2\farcm6$ at 115.271 GHz, the rest frequency of $^{12}$CO~$J$~=~1--0.
The observations were performed using the on-the-fly (OTF) mapping mode.
We used the 4 K cooled superconductor-insulator-superconductor (SIS) receiver in the front-end system. The system noise temperature shows $\sim 250$ K at the double-sideband (DSB). 
The back-end system is a digital-Fourier-transform spectrometer (DFS) with a 1 GHz bandwidth and 16384 channels.
The velocity bandwidth and resolution are 2600\kms and 0.16\kms at 115 GHz, respectively.
Observers checked the pointing accuracy within 15\arcsec\ by observing a variable carbon star IRC+10216 and the Sun.
The raw data, including the atmosphere, is converted to the antenna temperature scale ($T_A^*$) by applying the chopper-wheel method \citep{1976ApJS...30..247U,1981ApJ...250..341K}.
We daily observed IRAS 16293--2422 $(\alpha_{\rm J2000},\delta_{\rm J2000})=(16^\mathrm{h} 32^\mathrm{m} 23.3^\mathrm{s}, {-24\degr28\arcmin 39.\arcsec2})$ located in the Rho Ophiuchi star-forming region as a standard source. 
We calibrated $T_A^*$ to the main-beam temperature scale ($T_{\rm MB}$), assuming $T_{\rm MB}=18$ K \citep{2006AJ....131.2921R} as the peak intensity of IRAS 16293--2422.
The cube data is smoothed to be 180\arcsec\ spatial resolution using the kernel Gaussian function of 90\arcsec.
The grid size of the final data is $(l,b,v)=$(60\arcsec, 60\arcsec, 0.16\kms).
The root-mean-square（r.m.s.）noise level is $\sim 1.1$ K at $T_{\rm mb}$ scale.
The current status of the NANTEN2 radio telescope is described in \cite{2020SPIE11453E..3ZN}.

We also used the $^{12}$CO~$J$~=~1--0 survey data from the NANTEN 4 m telescope at {Las Campanas} observatory in Chile until 2004.
The detailed information of this data is described in \cite{2004ASPC..317...59M} and \cite{2010PASJ...62..557T}. The HPBW, grid spacing, and velocity resolution are {2\farcm6}, 4\arcmin, and 0.65\kms, respectively.
The grid size and r.m.s noise level of cube data used in this paper are $(l,b,v)=$(4\arcmin, 4\arcmin, 1\kms) and $\sim 0.2$ K at $T_{\rm mb}$ scale.

\subsection{The Mopra $^{12}$CO and $^{13}$CO~$J$~=~1--0 Galactic Plane survey data}
{To perform a detailed analysis focusing on RCW 106 East and RCW 106 West}, we utilized the high-resolution CO data obtained by the Mopra Southern Galactic Plane CO Survey \citep{2013PASA...30...44B,2015PASA...32...20B,2018PASA...35...29B,2023PASA...40...47C}\footnote{\url{https://mopracosurvey.wordpress.com}}.
The Mopra radio telescope, with a diameter of 22 m, is operated by the Australia Telescope National Facility (ATNF). The Mopra Southern Galactic plane survey was observed in the $^{12}$CO, $^{13}$CO, C$^{18}$O, and C$^{17}$O $J$~=~1--0 emission, carried out during the winter season from 2011 to 2018. 
{The observations were the Fast-On-The-Fly (FOTF) mode every 60\arcmin $\times$ 6\arcmin\ rectangle unit. The telescope performed orthogonal scans in the direction of Galactic longitude and latitude to reduce scanning artifacts.} {Part of the observations were conducted remotely from Nagoya University in Japan through an internet connection.}
The front end uses a Monolithic Microwave Integrated Circuit (MMIC) receiver covered with a 3 mm band. 
The back end is utilized as the UNSW Mopra Spectrometer (MOPS).
The HPBW of Mopra at 115 GHZ is 33\arcsec. The survey data is converted to the main-beam temperature scale using the relation of $T_{\rm MB} = T_{\rm A}^*/\eta$ with the extended beam efficiency of $\eta=0.55$ \citep{2005PASA...22...62L}.
The spatial and velocity resolution of the data release 4 (DR4) is 36\arcsec\ and 0.1\kms, respectively.
The grid size used in this paper is $(l,b,v)=$(30\arcsec, 30\arcsec, 1\kms).
The r.m.s noise levels are $\sim 0.8$  K in $^{12}$CO~$J$~=~1--0 and $\sim 0.3$ K in $^{13}$CO~$J$~=~1--0.

\subsection{The Herschel far-infrared archival data}
{To compare spatial distributions between molecular gas and massive star/cluster sources in the RCW 106 GMC complex}, we used the archival infrared image data obtained by Herschel Space Observatory\footnote{Herschel Space Observatory is an ESA space observatory with science instruments provided by European-led Principal Investigator consortia and with important participation from NASA.} \citep{2010A&A...518L...1P}.
The image data are taken from the VIALACTEA web page\footnote{\url{http://vialactea.iaps.inaf.it/vialactea/eng/index.php}} as a part of the Herschel infrared Galactic Plane Survey (Hi-GAL: \citealp{2010PASP..122..314M,2010A&A...518L.100M,2016A&A...591A.149M}). 
The far-infrared image data of 70 $\mu$m, 160 $\mu$m, and 350 $\mu$m are obtained by the Photodetector
Array Camera and Spectrometer (PACS: \citealp{2010A&A...518L...2P}) and Spectral and Photometric Imaging REceiver (SPIRE: \citealp{2010A&A...518L...3G}), respectively.
The spatial resolution of 70 $\mu$m, 160 $\mu$m, and 350 $\mu$m data are 6\arcsec, 12\arcsec, and 24\arcsec, respectively.
Table \ref{obs_para} summarizes the data properties in this paper.

\section{Results}
\subsection{CO spatial and velocity distributions of the RCW 106 GMC complex}

\begin{figure*}[h]
\begin{center} 
 \includegraphics[width=16cm]{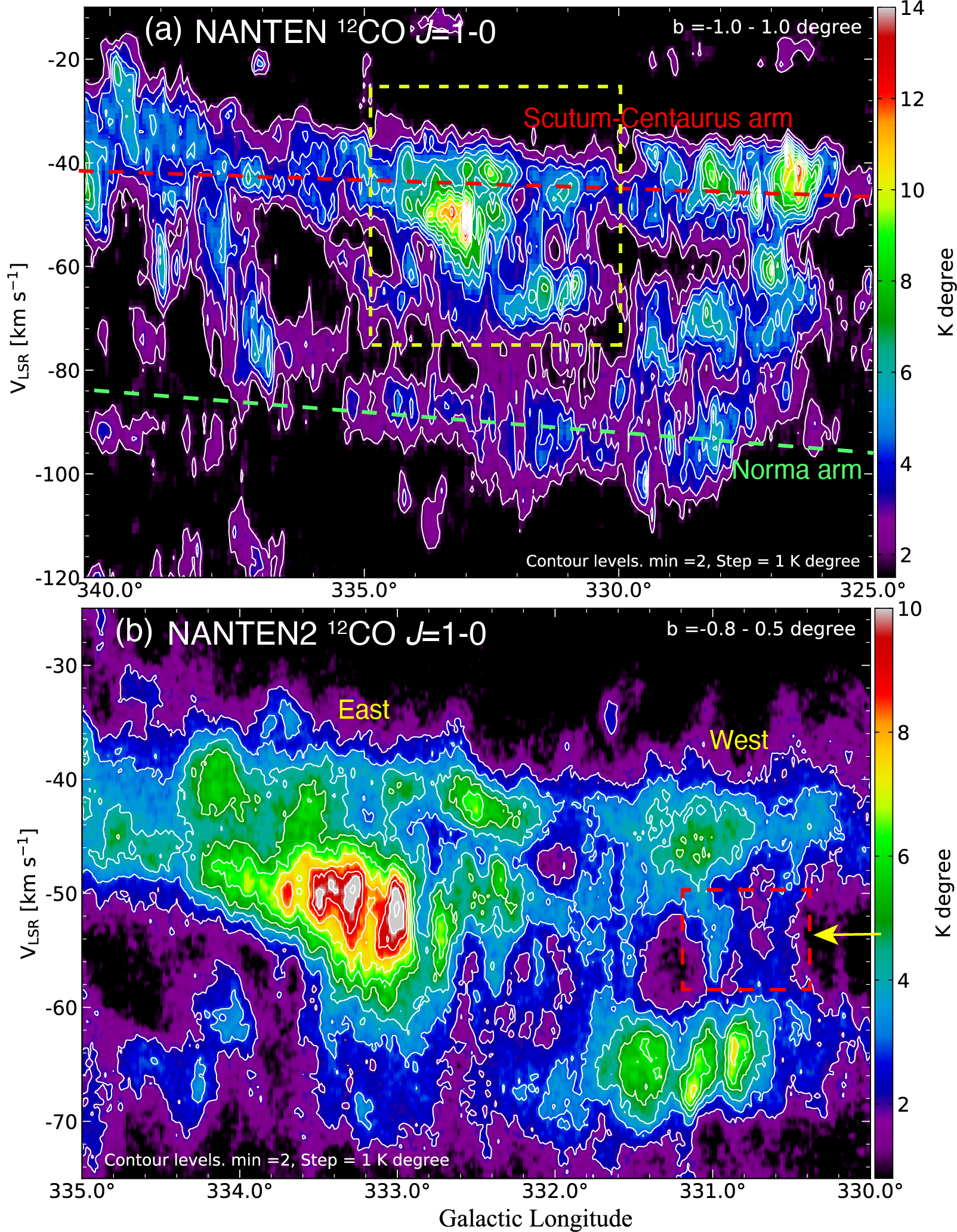}
\end{center}
\caption{(a) The large-scale longitude-velocity diagram of $^{12}$CO~$J$~=~1--0 obtained by NANTEN. The red and green dotted lines indicate the Scutum-Centaurus and Norma arm, respectively \citep{2017AstRv..13..113V,2016ApJ...823...76K}. {The integrated latitude range is from $-1.0$\degr\ to $1.0$\degr.} The yellow dotted square shows the area in panel (b). (b) {Detailed} longitude-velocity diagram {focusing on} the RCW 106 GMC complex of $^{12}$CO~$J$~=~1--0 obtained by NANTEN2. {The integrated latitude range is from $-0.8$\degr\ to $0.5$\degr.} {The red dotted rectangle and yellow arrow show the bridge feature at RCW 106 West.} The velocity space is smoothed in 9 channels by the hanning window. The lowest contour and interval are 2 K degree and 1 K degree, respectively.}
\label{NASCO_lv}
\end{figure*}

Figure \ref{colorimg}(d) shows the $^{12}$CO~$J$~=~1--0 integrated intensity map obtained by NANTEN2.
The molecular cloud distributes over {200} pc $\times\ 80$ pc ($l \times b$) at the Galactic plane from $l=330$\degr\ to $335$\degr.
The CO peak has $(l,b)\sim (333\fdg3,-0\fdg35)$, which corresponds to the active massive star-forming of G333 including G333.6--0.2, IRAS 16177--5018, IRAS 16172--5028,  G333.0--0.4, IRAS 16164--5046, and IRAS 16158--5055 (see Figure 3 of \citealp{2006MNRAS.367.1609B}).
The CO peaks coincide with the bright infrared regions at RCW 106 East and RCW 106 West, as shown in {Figures \ref{colorimg}(a) and (d)}.
Figure \ref{NASCO_lv}(a) presents the large-scale longitude-velocity diagram of $^{12}$CO~$J$~=~1--0 obtained by NANTEN.
The two velocity components of $-90$\kms and $-50$\kms at $l \sim 333 \degr$ correspond to the Norma arm and Scutum-Centaurus arm in the Milky Way, respectively \citep{2017AstRv..13..113V,2016ApJ...823...76K}.
The yellow dotted box shows the main components of the RCW 106 GMC complex.
Figure \ref{NASCO_lv}(b) shows the detailed longitude-velocity diagram of $^{12}$CO~$J$~=~1--0, focusing on the RCW 106 GMC complex obtained by NANTEN2.
{In this figure, we can find two velocity components of $\sim -68$\kms and $\sim -50$\kms in the GMC complex.} The CO gas has a peak at RCW 106 East. These two components are separated at RCW 106 West and connect on the velocity space with the bridge feature. 

\begin{figure*}[h]
\begin{center} 
\includegraphics[width=16cm]{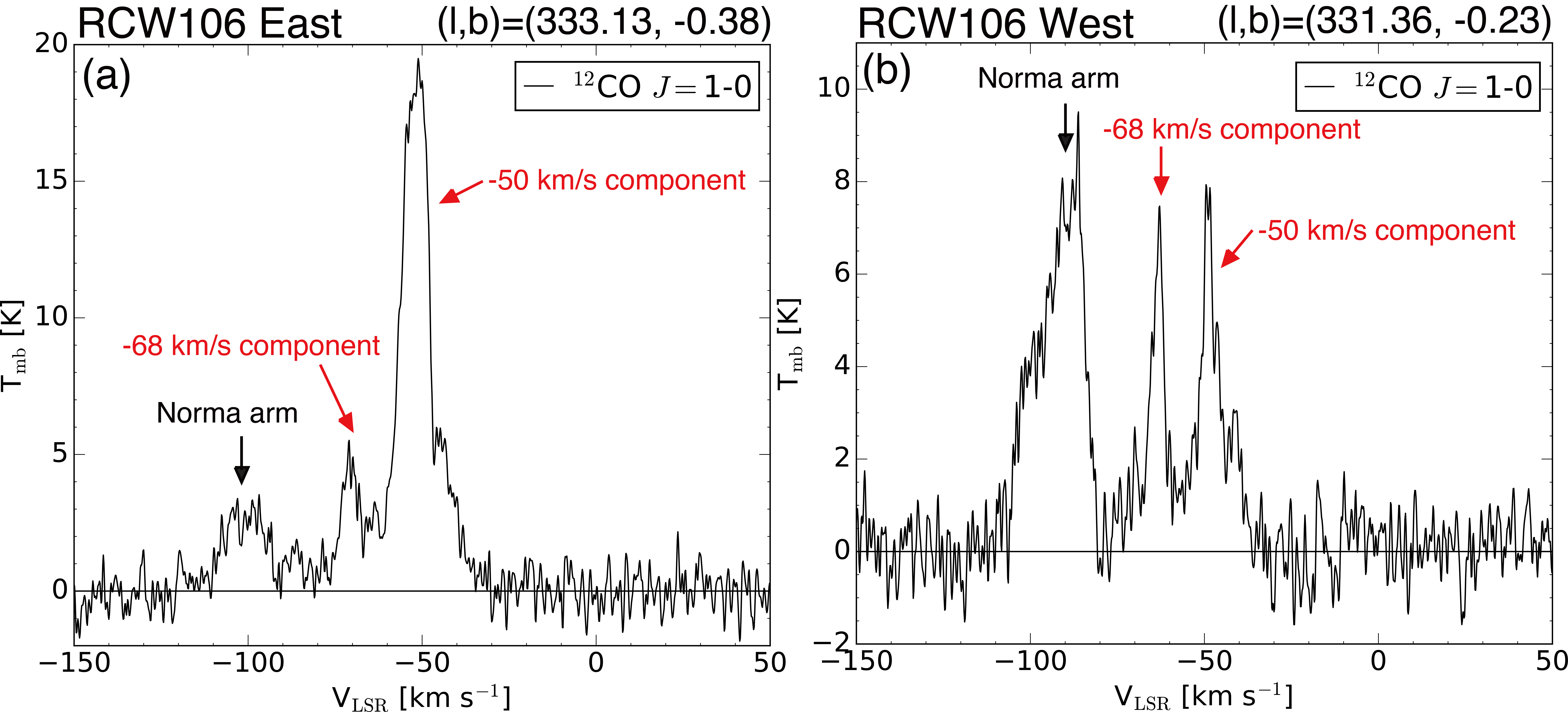}
\end{center} 
\caption{Spectra for $^{12}$CO~$J$~=~1--0 obtained at (a) RCW 106 East (G333), and (b) RCW 106 West with NANTEN2. The red arrows indicate the velocity components associated with the RCW 106 GMC complex. The black arrow shows the velocity component of the Norma arm. The spectra are smoothed to the velocity domain of 9 channels by the hanning window. The position of each spectrum is shown in the upper right of each panel and {the orange cross marks in Figures \ref{2GMCs}a and \ref{2GMCs}b.}}
\label{spec}
\end{figure*}

{Figures \ref{spec} (a) and (b)} show the $^{12}$CO~$J$~=~1--0 spectra in RCW 106 East and RCW 106 West, respectively. {The positions of the spectra are shown as the orange cross marks in Figures \ref{2GMCs}a and \ref{2GMCs}b.} Previous studies indicate that the velocity range from $-110$\kms to $-90$\kms is associated with the Norma arm with a distance of $\sim$ 4.6--7.5 kpc and the background component toward RCW 106 \citep{2013ApJ...774...38M,2015ApJ...812....7N}.
The velocity range from $-80$\kms to $-30$\kms is associated with RCW 106 located on the Scutum-Centaurus arm and is composed of two velocity components, which are from {$-75$\kms to $-60$\kms and from $-60$\kms to $-45$\kms.
Hereafter, we call these components ``$-68$\kms GMC" and ``$-50$\kms GMC"}.

Figure \ref{ch} shows the velocity channel map of $^{12}$CO~$J$~=~1--0 from $-80$\kms to $-30$\kms. 
Molecular gas distributes from $-70$\kms to $-35$\kms.
We can find the two filamentary structures in RCW 106 East and West from $-53$\kms to $-49$\kms.
Their lengths and widths are $\sim 150$ pc and $\sim 20$ pc at a distance of 3.6 kpc.
Previous studies identified these filaments as the giant molecular filament of GMF 335.6−-333.6a and GMF 335.6−-333.6b \citep{2016A&A...590A.131A,2018ApJ...864..153Z,2019A&A...622A..52Z}.

\begin{figure*}[h]
\begin{center} 
 \includegraphics[width=19cm]{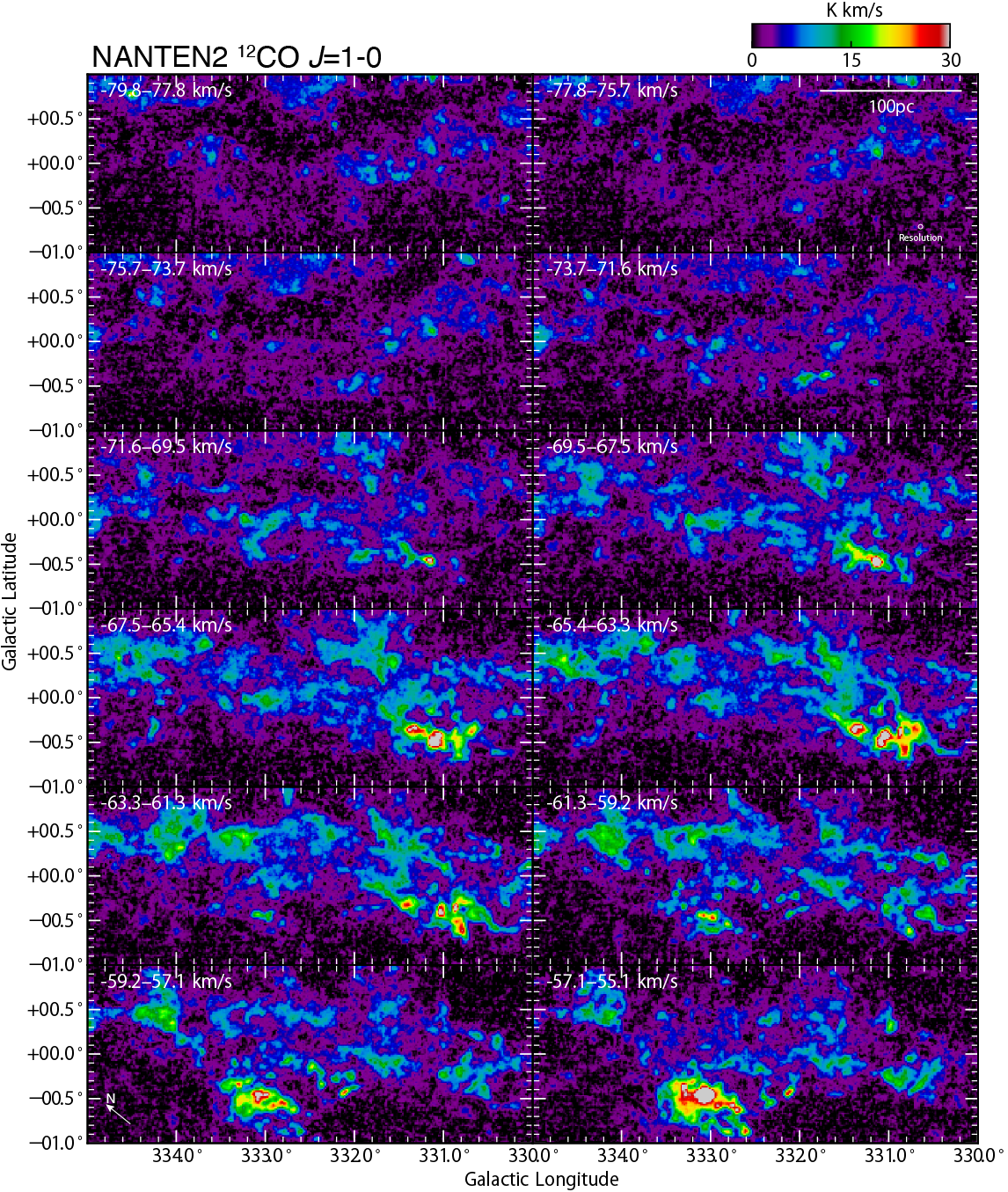}
\end{center}
\caption{The $^{12}$CO~$J$~=~1--0 velocity channel map obtained by NANTEN2. The cube data is smoothed to the velocity domain of 9 channels by the hanning window. Red dotted rectangles show giant molecular filaments of GMF 335.6−333.6a and GMF 335.6−333.6b.}
\label{ch}
\end{figure*}
\addtocounter{figure}{-1}

\begin{figure*}[h]
\begin{center} 
 \includegraphics[width=19cm]{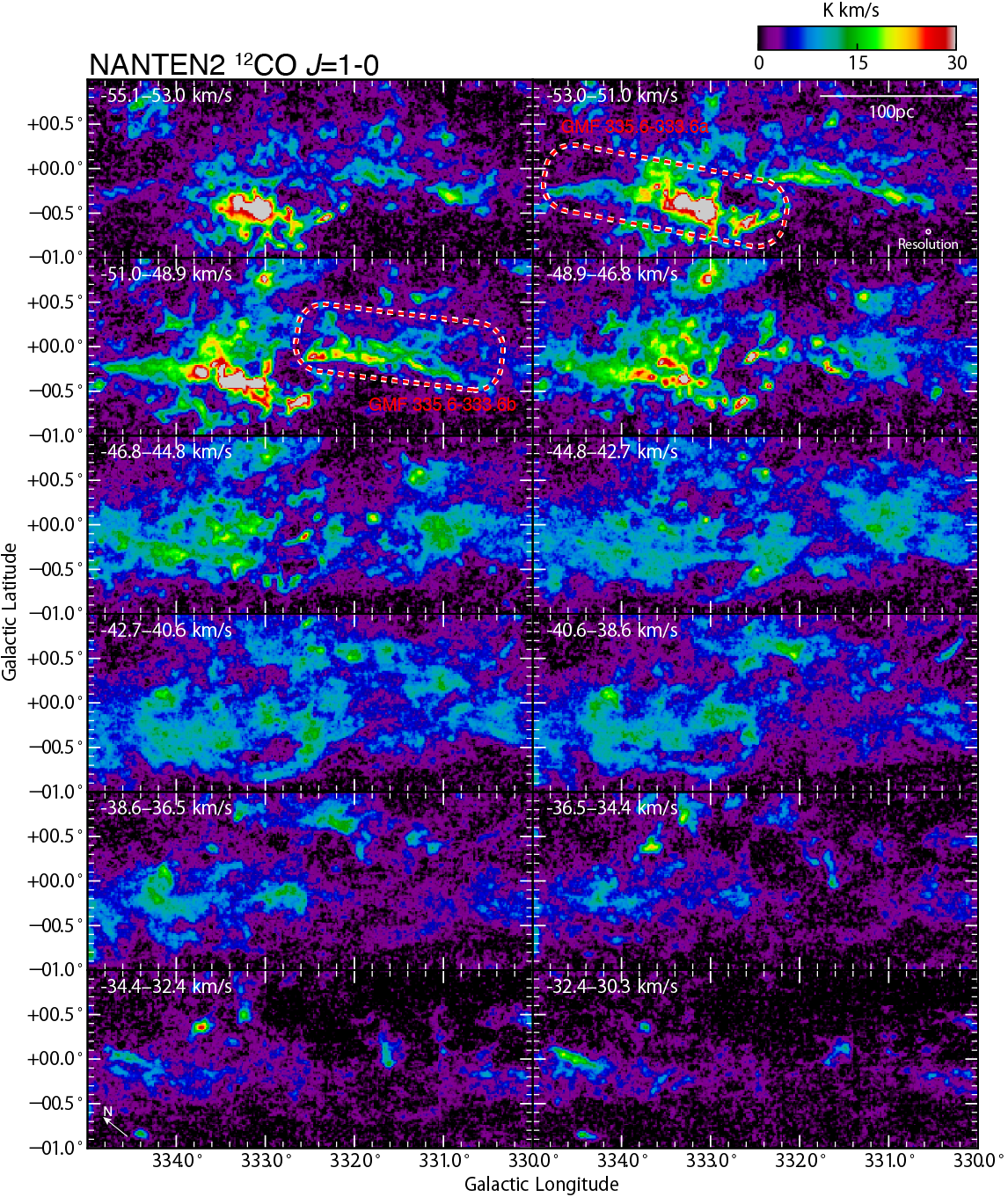}
\end{center}
\caption{Continued. }
\label{ch}
\end{figure*}

\subsection{Two giant molecular clouds}
\begin{figure*}[h]
 \includegraphics[width=17cm]{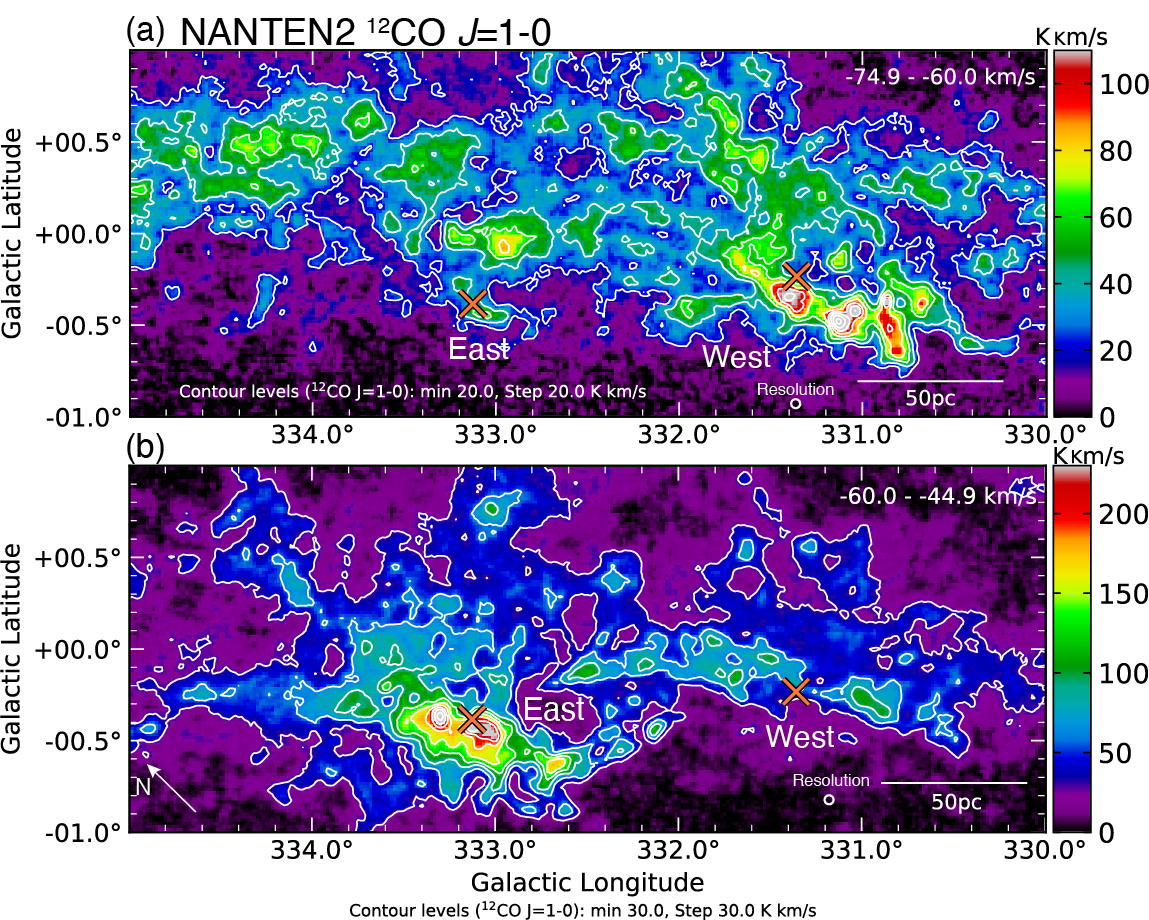}
\caption{The $^{12}$CO~$J$~=~1--0 integrated intensity maps of (a) the $-68$\kms GMC and (b) $-50$\kms GMC obtained by NANTEN2. The integrated velocity ranges of (a) and (b) are from $-75$\kms to $-60$\kms and from $-60$\kms to {$-45$\kms}\, respectively. The lowest contour and interval of the panel (a) are 20 K\kms and 20 K\kms, respectively. The lowest contour and interval of the panel (b) are {30 K\kms and 30 K\kms}, respectively. {The orange cross marks at East and West show the position of spectra in Figures \ref{spec}a and \ref{spec}b, respectively.}}
\label{2GMCs}
\end{figure*}

\begin{figure*}[h]
 \includegraphics[width=17cm]{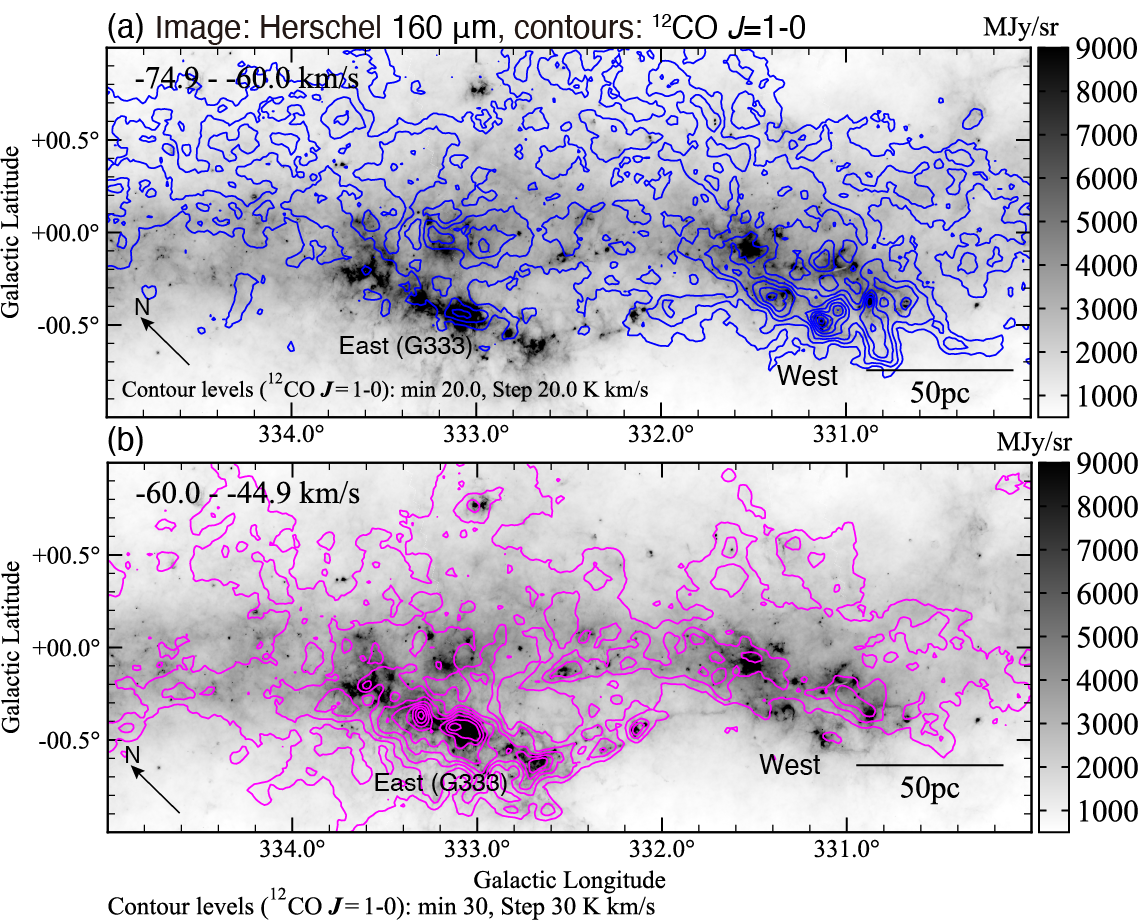}
\caption{The $^{12}$CO~$J$~=~1--0 integrated intensity maps of (a) the $-68$\kms GMC and (b) $-50$\kms GMC {obtained by NANTEN2} superposed on the Herschel 160 $\mu$m continuum image \citep{2010PASP..122..314M}. The lowest contour and interval of the panel (a) is 20 K\kms. The lowest contour level and interval of the panel (b) is {30 K\kms}.}
\label{infra}
\end{figure*}

{Figures \ref{2GMCs}(a) and (b)} display the $^{12}$CO~$J$~=~1--0 integrated intensity maps of the $-68$\kms and $-50$\kms GMC.
The $-68$\kms GMC exhibits a peak at RCW 106 West and a filamentary structure from North to South.
The $-50$\kms GMC elongates along the Galactic plane and peaks at RCW 106 East.
We calculated the H$_2$ column densities and molecular masses {within the range above the $5\sigma$ noise level with the $^{12}$CO integrated intensity}, assuming the CO-to-H2 conversion factor of $2 \times 10^{20}\ {\rm cm^{-2}\ (K\ km\ s^{-1})^{-1}}$ \citep{2013ARA&A..51..207B}.
{The peak H$_2$ column densities of the $-68$\kms GMC and $-50$\kms GMC are $3.3 \times 10^{22}$ cm$^{-2}$ and $5.7 \times 10^{22}$ cm$^{-2}$, respectively.
The total molecular mass of $-68$\kms GMC and $-50$\kms GMC are $\sim 4.0 \times 10^6\ M_{\odot}$ and $\sim 5.5 \times 10^6\ M_{\odot}$, respectively.}
In the appendix, we present detailed procedures for derived physical properties of molecular clouds.
{Figures \ref{infra}(a) and (b)} show the $-68$\kms and $-50$\kms GMC superposed on the Herschel 160 $\mu$m dust continuum image \citep{2010PASP..122..314M}.
We point out that peaks of molecular gas coincide with bright cold dust emission, as shown in the Herschel infrared image at RCW 106 East and RCW 106 West.
{These results indicate that the molecular clouds with a velocity difference of $\sim 20$\kms are likely to be physically associated with the active star-forming regions in the RCW 106 GMC complex.
Based on the large-scale observational results, we performed a detailed analysis focusing on RCW 106 East and RCW 106 West to reveal the relationship between two GMCs and massive star/cluster formation in the RCW 106 GMC complex in the next section.}


\clearpage
\subsection{RCW 106 East (G333)}

\begin{figure*}[h]
\begin{center} 
\includegraphics[width=17cm]{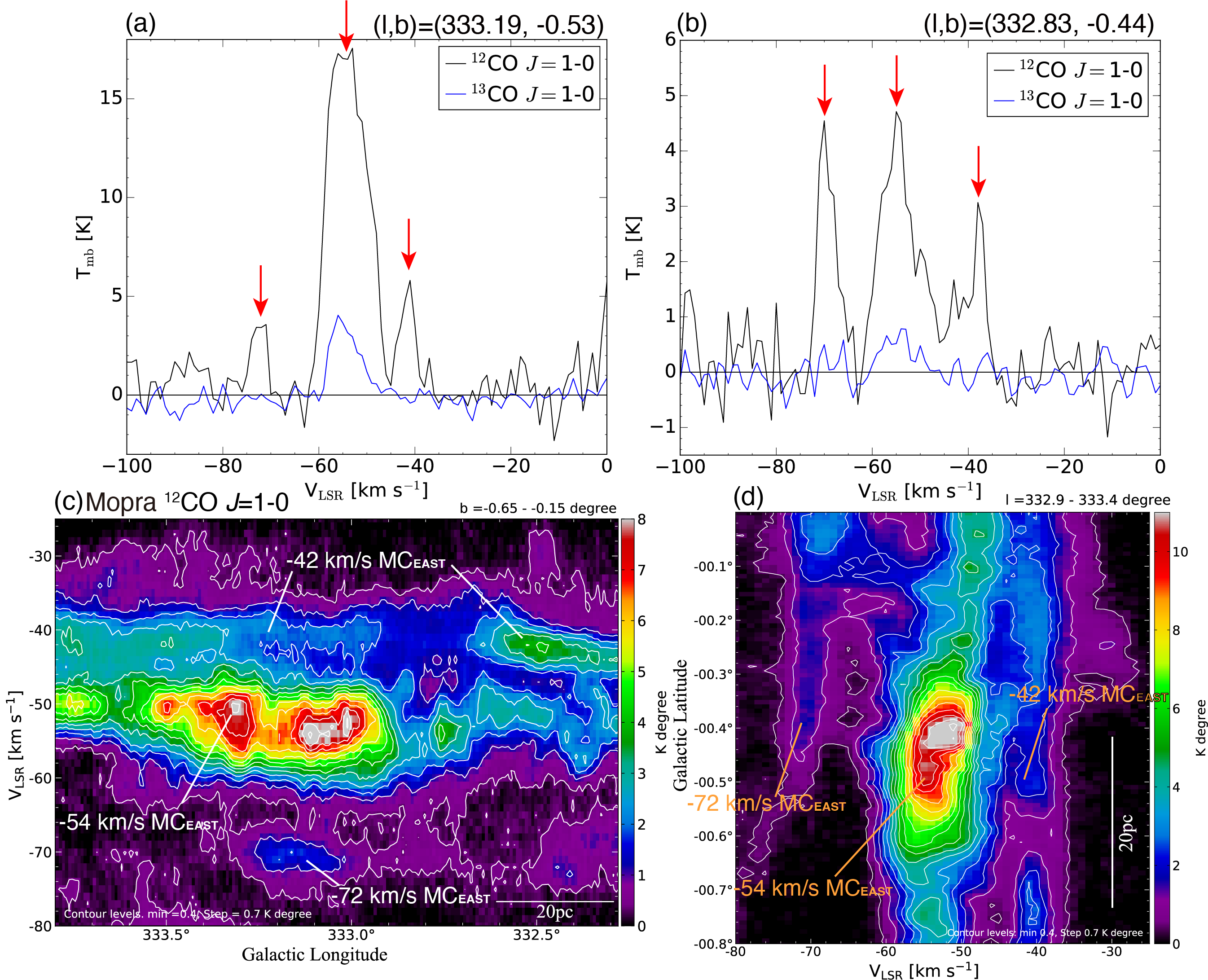}
\end{center}
\caption{The CO~$J$~=~1--0 spectra in RCW 106 East at (a) $(l,b)=(333\fdg19, -0\fdg53)$ and (b) $(l,b)=(332\fdg83, -0\fdg44)$ obtained by Mopra. Black and blue spectra show $^{12}$CO and $^{13}$CO~$J$~=~1--0 emission, respectively. Red arrows present the three velocity components of $-72$ km s$^{-1}$, $-54$ km s$^{-1}$, and $-42$ km s$^{-1}$. {The position of spectra in the panel (a) and (b) is the green circles of A and B in Figure \ref{overlay}a, respectively.} (c) The $^{12}$CO~$J$~=~1--0 longitude-velocity diagram in RCW 106 East. The integrated latitude range is from $-0\fdg65$ to $-0\fdg15$. (d) The $^{12}$CO~$J$~=~1--0 latitude-velocity diagram in RCW 106 East. The integrated longitude range is from $332\fdg9$ to $333\fdg4$. The lowest contour levels and intervals are 0.4 K degree and 0.7 K degree, respectively.}
\label{eastpv}
\end{figure*}
{Figures \ref{eastpv}(a) and (b)} show the CO spectra at $(l,b)=(333\fdg19, -0\fdg53)$ and $(l,b)=(332\fdg83, -0\fdg44)$ in RCW 106 East obtained by Mopra, respectively. {The positions of the spectra are shown as the green circles of A and B in Figure \ref{overlay}a.}
We can find that three velocity components of $-72$ km s$^{-1}$, $-54$ km s$^{-1}$, and $-42$ km s$^{-1}$ in $^{12}$CO~$J$~=~1--0 shown by red arrows. 
{Hereafter, we call these molecular clouds (MCs) the $-72$ km s$^{-1}$ MC$_{\rm EAST}$, $-54$ km s$^{-1}$ MC$_{\rm EAST}$, and $-42$ km s$^{-1}$ MC$_{\rm EAST}$.}
{Figures \ref{eastpv}(c) and (d)} present the longitude-velocity and latitude-velocity diagram at RCW 106 East, respectively. 
These MCs connect on the velocity space in the position-velocity diagrams. 


\begin{figure*}[h]
\begin{center} 
\includegraphics[width=13cm]{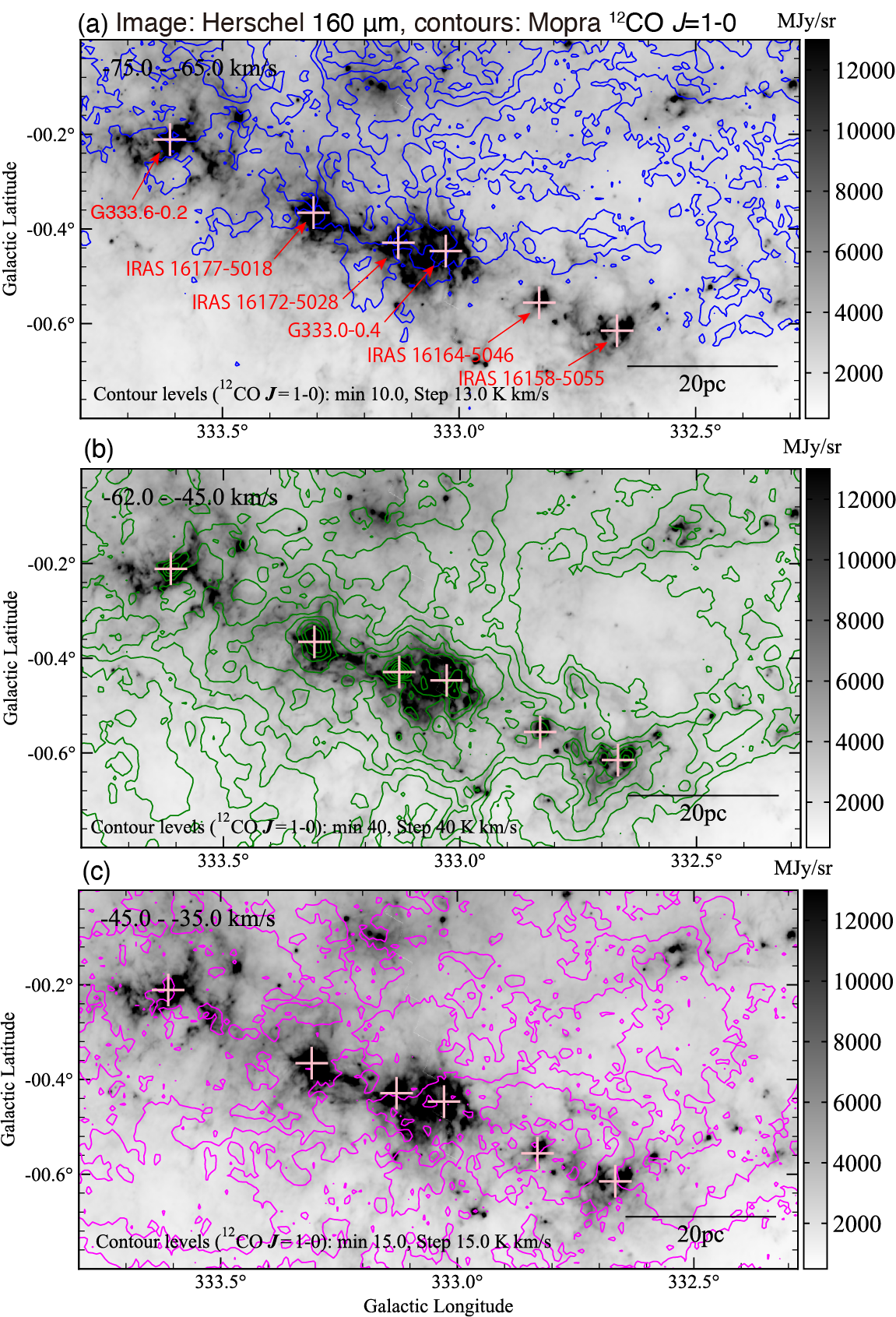}
\end{center}
\caption{The $^{12}$CO~$J$~=~1--0 integrated intensity maps of the three velocity components in RCW 106 East (G333) superposed on the Herschel $160\ \mu$m continuum image \citep{2010PASP..122..314M} for the (a) $-72$ km s$^{-1}$ {MC$_{\rm EAST}$}, (b) $-54$ km s$^{-1}$ {MC$_{\rm EAST}$}, and (c) $-42$ km s$^{-1}$ {MC$_{\rm EAST}$}. The data are smoothed to be 50\arcsec\ using a kernel Gaussian function. The cross marks are positions of the massive star-forming regions in RCW 106 East \citep{2006MNRAS.367.1609B}. 
The integrated velocity range of $-72$ km s$^{-1}$\ {MC$_{\rm EAST}$}, $-54$ km s$^{-1}$\ {MC$_{\rm EAST}$}, and $-42$ km s$^{-1}$\ {MC$_{\rm EAST}$} is from $-75$\kms to $-65$\kms, from $-62$\kms to $-45$\kms, and from $-45$\kms to $-35$\kms, respectively.
The lowest contour level and interval are 10 K\kms and 13 K\kms in the panel (a), 40 K\kms and 40 K\kms in the panel (b), 15 K\kms and 15 K\kms in the panel (c).}
\label{east}
\end{figure*}
{Figures \ref{east}(a),(b), and (c)} show the $-72$ km s$^{-1}$ {MC$_{\rm EAST}$}, $-54$ km s$^{-1}$ {MC$_{\rm EAST}$}, and $-42$ km s$^{-1}$ {MC$_{\rm EAST}$} superposed on the Herschel 160 $\mu$m continuum image, respectively.
The $-72$ km s$^{-1}$ {MC$_{\rm EAST}$} distributes at the massive star-forming region of {G333.6}--0.2, IRAS 16177--5018, IRAS 16172--5028, and G333.0--0.4.
The $-54$ km s$^{-1}$ {MC$_{\rm EAST}$} is the main intensity component, and molecular gas coincides with infrared peaks.
The $-42$ km s$^{-1}$ {MC$_{\rm EAST}$} distributes throughout RCW 106 East.
We point out that {$-72$ km s$^{-1}$ {MC$_{\rm EAST}$} and $-54$ km s$^{-1}$ {MC$_{\rm EAST}$}} overlap the massive star-forming regions of {G333.6}--0.2, IRAS 16172--5028, and G333.0--0.4.

\subsection{RCW 106 West}
\begin{figure*}[h]
\begin{center} 
\includegraphics[width=17cm]{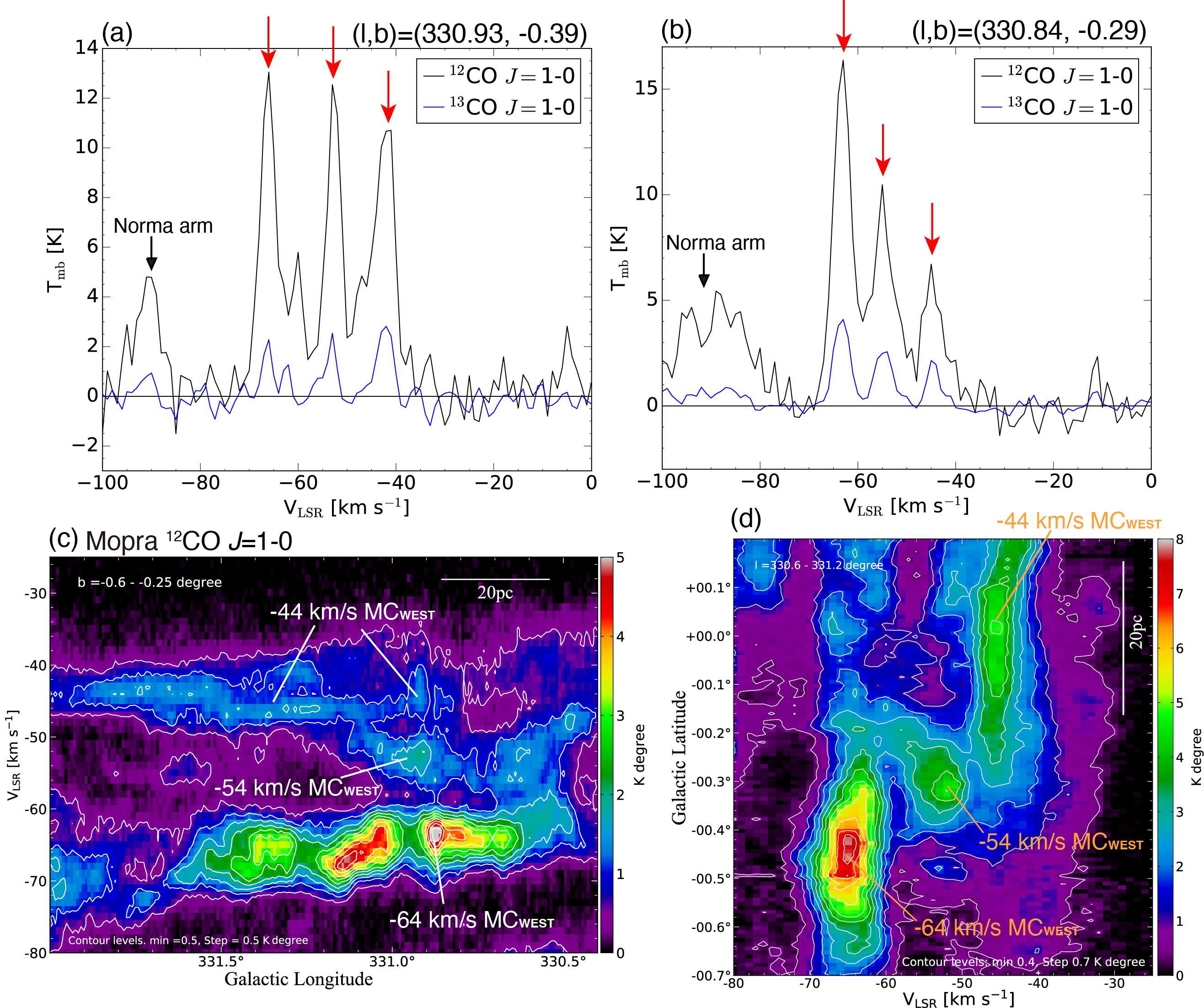}
\end{center}
\caption{The CO~$J$~=~1--0 spectra in RCW 106 West at (a) $(l,b)=(330\fdg93, -0\fdg39)$ and (b) $(l,b)=(330\fdg84, -0\fdg29)$ obtained by Mopra. Black and blue spectra show $^{12}$CO and $^{13}$CO~$J$~=~1--0 emission, respectively. Red arrows present the three velocity components of $-64$ km s$^{-1}$, $-54$ km s$^{-1}$, and $-44$ km s$^{-1}$. The black arrows show the velocity components of the Norma arm. {The position of spectra in the panel (a) and (b) is the green circles of C and D in Figure \ref{overlay}b, respectively.} (c) The $^{12}$CO~$J$~=~1--0 longitude-velocity diagram in RCW 106 West. The integrated latitude range is from $-0\fdg60$ to $-0\fdg25$. (d) The $^{12}$CO~$J$~=~1--0 latitude-velocity diagram in RCW 106 West. The integrated longitude range is from $330\fdg6$ to $331\fdg2$. The lowest contour levels and intervals are 0.5 K degree in the panel (c). The lowest contour levels and intervals are 0.4 K degree and 0.7 K degree in the panel (d), respectively.}
\label{westpv}
\end{figure*}
{Figures \ref{westpv}(a) and (b)} show the CO spectra at $(l,b)=(330\fdg93, -0\fdg39)$ and $(l,b)=(330\fdg84, -0\fdg29)$ in RCW 106 West. {The positions of the spectra are shown as the green circles of C and D in Figure \ref{overlay}b.}
We can find the three velocity components of $-64$ km s$^{-1}$, $-54$ km s$^{-1}$, and $-44$ km s$^{-1}$ in $^{12}$CO~$J$~=~1--0 presented by red arrows. 
{Hereafter, we call these MCs the $-64$ km s$^{-1}$ MC$_{\rm WEST}$, $-54$ km s$^{-1}$ MC$_{\rm WEST}$, and $-44$ km s$^{-1}$ MC$_{\rm WEST}$.} 
The black arrow shows the Norma arm component.
{Figures \ref{westpv}(c) and (d)} demonstrate the longitude-velocity and latitude-velocity diagrams, respectively.
These MCs in RCW 106 West also connect at $l \sim 330\fdg85$ and $b \sim -0\fdg30$ on the velocity space.


\begin{figure*}[h]
\begin{center} 
\includegraphics[width=13cm]{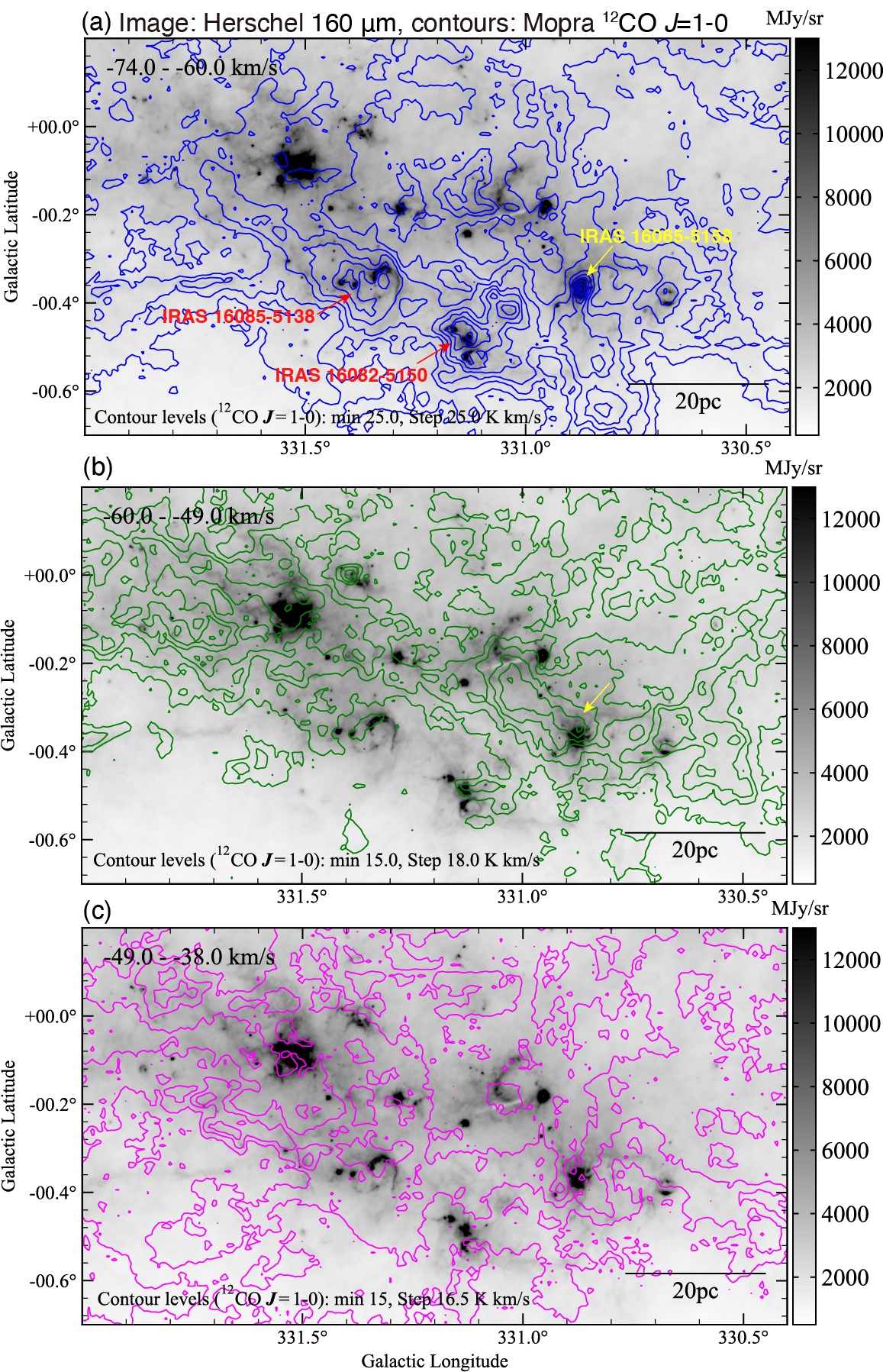}
\end{center}
\caption{The $^{12}$CO~$J$~=~1--0 integrated intensity maps of the three velocity components in RCW 106 West superposed on the Herschel $160\ \mu$m continuum image \citep{2010PASP..122..314M} for the (a) $-64$ km s$^{-1}$\ {MC$_{\rm WEST}$}, (b) $-54$ km s$^{-1}$\ {MC$_{\rm WEST}$}, and (c) $-44$ km s$^{-1}$ {MC$_{\rm WEST}$}. The data are smoothed to be 50\arcsec\ using a kernel Gaussian function. The integrated velocity range of $-64$ km s$^{-1}$\ {MC$_{\rm WEST}$}, $-54$ km s$^{-1}$\ {MC$_{\rm WEST}$}, and $-44$ km s$^{-1}$\ {MC$_{\rm WEST}$} is from $-74$\kms to $-60$\kms, from $-60$\kms to $-49$\kms, and from $-49$\kms to $-38$\kms, respectively. The lowest contour level and interval are 25 K\kms and 25 K\kms in the panel (a), 15 K\kms and 18 K\kms in the panel (b), 15 K\kms and 16.5 K\kms in the panel (c).}
\label{west}
\end{figure*}
{Figures \ref{west}(a),(b), and (c)} show the $-64$ km s$^{-1}$ {MC$_{\rm WEST}$}, $-54$ km s$^{-1}$ {MC$_{\rm WEST}$}, and $-44$ km s$^{-1}$ {MC$_{\rm WEST}$} overlaid on the Herschel 160 $\mu$m continuum image.
The main component of $-64$\kms {MC$_{\rm WEST}$} has CO peaks corresponding to bright infrared emission at the massive star-forming regions of IRAS 16085--5138, IRAS 16082--5150, and IRAS 16065--5158.
In particular, {$-64$ km s$^{-1}$ {MC$_{\rm WEST}$} and $-54$ km s$^{-1}$ {MC$_{\rm WEST}$}} are overlapped at IRAS 16065--5158 and coincide molecular gas with the infrared peak.

\begin{figure*}[h]
\begin{center} 
\includegraphics[width=18cm]{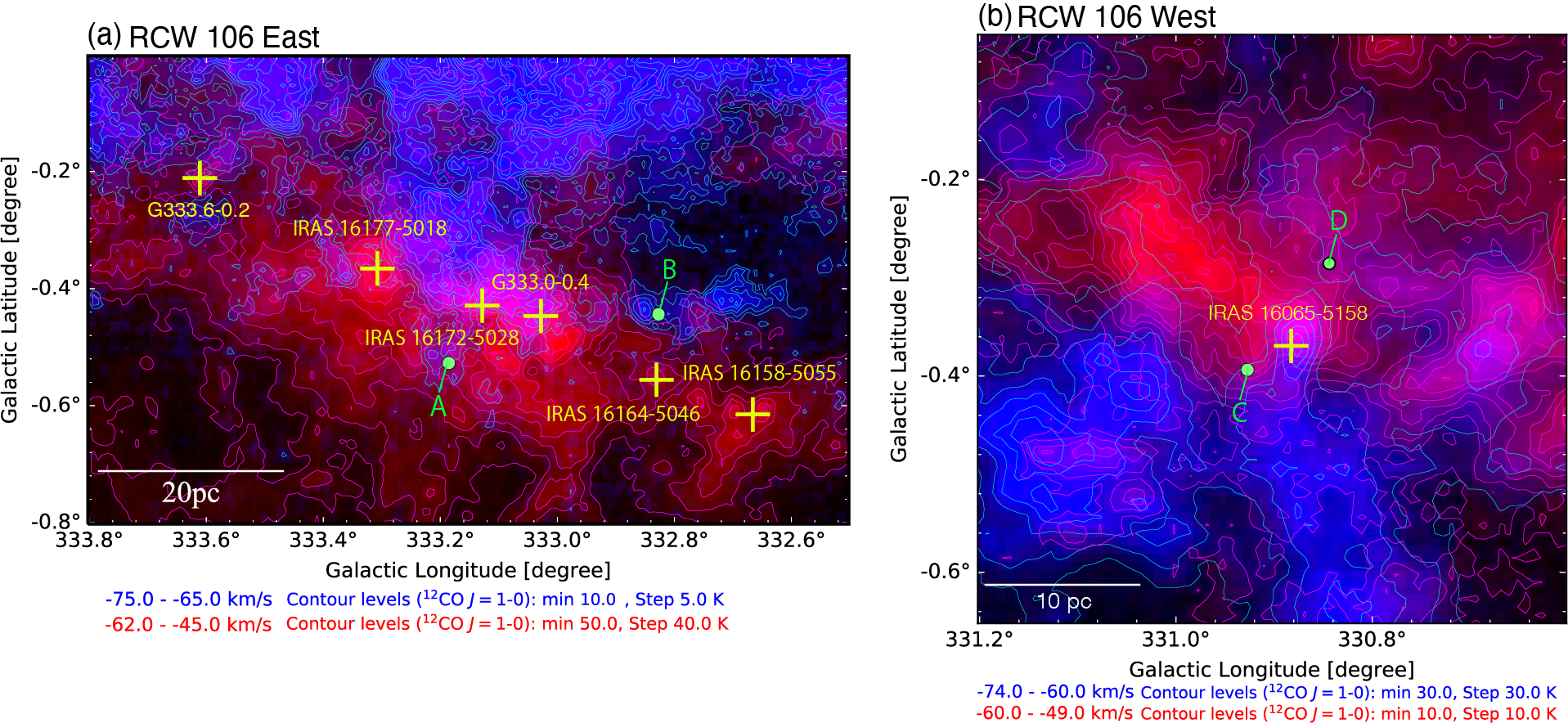}
\end{center}
\caption{{The two-color composite image of the $^{12}$CO~$J$~=~1--0 integrated intensity distributions obtained by Mopra. (a) The blue and red image represent $-72$\kms {MC$_{\rm EAST}$} and $-54$\kms {MC$_{\rm EAST}$}, respectively. The integrated velocity ranges and yellow crosses are the same as Figure \ref{east}. The lowest contour levels and intervals are 10 K\kms and 5 K\kms in $-72$\kms {MC$_{\rm EAST}$}, 50 K\kms and 40 K\kms in $-54$\kms {MC$_{\rm EAST}$}, respectively. 
(b) The blue and red image represent $-64$\kms {MC$_{\rm WEST}$} and $-54$\kms {MC$_{\rm WEST}$ obtained by Mopra}, respectively. The integrated velocity ranges and yellow cross are the same as Figure \ref{west}. The lowest contour levels and intervals are 30 K\kms and 30 K\kms in $-64$\kms {MC$_{\rm WEST}$}, 10 K\kms and 10 K\kms in $-54$\kms {MC$_{\rm WEST}$}, respectively. The green circles of A, B, C, and D show the position of spectra in Figure \ref{eastpv}a, Figure \ref{eastpv}b, Figure \ref{westpv}a, and Figure \ref{westpv}b, respectively.}}
\label{overlay}
\end{figure*}

We estimated the physical parameters of the MCs in RCW 106 East and RCW 106 West with data points above the $5\sigma$ noise levels in $^{12}$CO and $^{13}$CO~$J$~=~1--0.
The parameters are excitation temperature ($T_{\rm ex}$), optical depth ($\tau_{13}$), H$_2$ column density, and total molecular mass calculated from the $^{12}$CO and $^{13}$CO~$J$~=~1--0 intensity.
Detailed procedures are presented in Appendix A. We summarize the results in Table \ref{param}.

\subsection{{Physical association of two MCs in RCW 106 East and West}}
{The $-42$ km s$^{-1}$ MC$_{\rm EAST}$ and $-44$ km s$^{-1}$\ {MC$_{\rm WEST}$}
 spatially extended distributions and no clumpy structures (see Figures \ref{ch}, \ref{east}c, and \ref{west}c).
\cite{2005A&A...429..497R} shows that these velocity components are associated with the young supernova remnant RCW 103 and Wolf-Rayet nebula RCW 104 as more evolved regions of RCW 106 from a velocity-resolved deep H${\alpha}$ survey. We point out that $-42$ km s$^{-1}$ MC$_{\rm EAST}$ and $-44$ km s$^{-1}$\ {MC$_{\rm WEST}$} might be diffuse foreground components at the front face of the Scutum-Centaurus arm.}
To summarize our result, focusing on the RCW 106 East and West regions internal the RCW 106 GMC complex, these regions comprise {two} MCs with {$\sim 10$\kms} differences associated with massive star-forming regions.

\begin{table*}[h]
\caption{Physical properties of molecular clouds in RCW 106 East and West.}
\begin{center}
\begin{tabular}{ccccccccccccccccc}
\hline
\multicolumn{1}{c}{Name} &$T_{\rm ex}$ & $\tau_{13}$ &$N^{12}_{\rm X\ max}$ & $N^{12}_{\rm X\ mean}$ & $N^{13}_{\rm LTE\ max}$ & $N^{13}_{\rm LTE\ mean}$& $M^{12}_{\rm X}$ & $M_{\rm LTE}^{13}$\\
&[K] & &  [cm$^{-2}$] & [cm$^{-2}$] & [cm$^{-2}$] & [cm$^{-2}$] & [$M_{\odot}$] & [$M_{\odot}$]  \\
(1) & (2) & (3) &(4)& (5) & (6) & (7) & (8) & (9)  \\
\hline\hline
RCW 106 East (total)  & 16 & 0.30  & $ 1.1\times 10^{23}$ &$ 2.1\times 10^{22}$& $2.8 \times 10^{23}$& $1.5 \times 10^{22}$ &$2.2 \times 10^{6}$ & $1.2 \times 10^{6}$ \\
$-72$ km s$^{-1}$ {MC$_{\rm EAST}$} & 15 & 0.33 & $1.4\times 10^{22}$ & $8.0 \times 10^{20}$ & $1.1 \times 10^{22}$& $2.4 \times 10^{21}$ &$ 8.6\times 10^{4}$ & $1.1 \times 10^{4}$ \\
$-54$ km s$^{-1}$ {MC$_{\rm EAST}$}  & 16 &  0.30 & $9.5 \times 10^{22}$ &$ 1.5\times 10^{22}$ & $2.7 \times 10^{23}$& $1.5 \times 10^{22}$ &$1.6 \times 10^{6}$ & $9.9 \times 10^{5}$ \\
$-42$ km s$^{-1}$ {MC$_{\rm EAST}$}  & 18 & 0.38 & $1.6 \times 10^{22}$ & $5.2 \times 10^{21}$ & $2.7 \times 10^{22}$& $2.6 \times 10^{21}$ &$5.6 \times 10^{5}$ & $1.1 \times 10^{5}$ \\
\hline
RCW 106 West (total)  & 12 & 0.32  & $6.6 \times 10^{22}$ & $1.2 \times 10^{22}$ & $ 9.7\times 10^{22}$& $5.9 \times 10^{21}$ &$ 1.6\times 10^{6}$ & $4.6 \times 10^{5}$ \\
$-64$ km s$^{-1}$ {MC$_{\rm WEST}$} & 12 & 0.29 & $4.1 \times 10^{22}$ & $5.5 \times 10^{21}$ & $8.4 \times 10^{22}$& $6.5 \times 10^{21}$ & $7.0 \times 10^{5}$ & $2.5 \times 10^{5}$ \\
$-54$ km s$^{-1}$ {MC$_{\rm WEST}$} &11  & 0.31 & $2.3 \times 10^{22}$  & $3.2 \times 10^{21}$ & $2.5 \times 10^{22}$ & $3.4 \times 10^{21}$& $ 4.1 \times 10^{5}$ &$ 9.1\times 10^{4}$  \\
$-44$ km s$^{-1}$ {MC$_{\rm WEST}$}  & 10 & 0.35 & $1.6\times 10^{22}$ & $3.7 \times 10^{21}$ & $2.1 \times 10^{22}$& $2.4 \times 10^{21}$ &$4.7 \times 10^{5}$ & $1.1 \times 10^{5}$ \\
\hline
\hline
\end{tabular}
  \vspace{3pt}
\label{param}\\
{\raggedright Notes. --- Columns: (1) The cloud name. (2) The mean excitation temperature obtained by the $^{12}$CO peak intensity. (3) The mean optical depth obtained by $^{13}$CO. (4) The peak H$_2$ column density derived from $^{12}$CO assuming the $X_{\rm CO}$ factor. (5) The mean H$_2$ column density calculated from $^{12}$CO assuming the $X_{\rm CO}$ factor. (6) The peak H$_2$ column density derived from $^{13}$CO assuming LTE. (7) The mean H$_2$ column density derived from $^{13}$CO assuming LTE. (8) The total molecular mass estimated from the H$_2$ column density assuming the $X_{\rm CO}$ factor. (9) The total molecular mass estimated from the H$_2$ column density assuming LTE.  The physical properties are derived above the $5\sigma$ noise level in the $^{12}$CO and $^{13}$CO cube data. \par}
\end{center}
\end{table*}

\section{{Discussion}}

\subsection{{The scenario of massive stars/cluster formation in the RCW 106 GMC complex}}
\begin{figure*}[h]
\begin{center} 
\includegraphics[width=18cm]{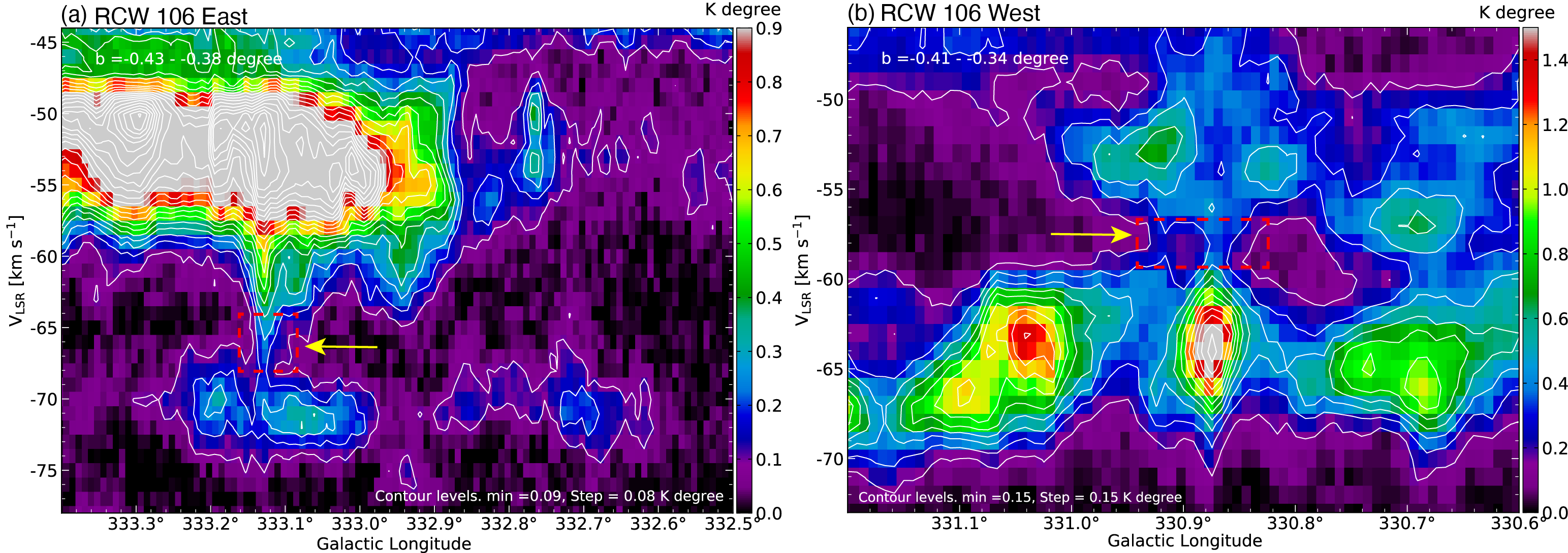}
\end{center}
\caption{{Detailed $^{12}$CO $J=$1--0 longitude-velocity diagrams of (a) RCW 106 East and (b) RCW 106 West. The integrated latitude ranges of (a) and (b) are from $-0.43\degr$ to $-0.38\degr$ and from $-0.41\degr$ to $-0.34\degr$, respectively. The lowest contour levels of (a) and (b) are 0.09~K~degree and 0.15~K~degree. The contour intervals of (a) and (b) are 0.08~K~degree and 0.15~K~degree. The red-dotted rectangles and yellow arrows show the bridge features connecting two velocity components.}}
\label{bridge}
\end{figure*}

{Our detailed analysis of RCW 106 East and RCW 106 West indicates that the {two} MCs having a velocity difference
of $\sim 10$\kms are physically associated with massive star-forming regions in the RCW 106 GMC complex.}
Figures \ref{overlay}(a) and \ref{overlay}(b) show the overlaid spatial distributions of {two} MCs focusing on RCW 106 East and RCW 106 West, respectively. These MCs overlap around massive star-forming regions of G333.6-0.2, IRAS 16172-5028, G333.0-0.4 in RCW 106 East, and IRAS 16065-5158 in RCW 106 West. This observational feature is consistent with the signature of cloud-cloud collisions reported by previous studies of Galactic min-starburst region NGC 6334 (see Figure 8 in \citealp{2018PASJ...70S..41F}).
{Figures \ref{bridge} (a) and (b) show the position-velocity diagrams focusing on the massive star-forming regions in RCW 106 East and West, respectively. 
We can find the bridge features connecting two CO velocity components on the velocity space. If two molecular clouds collide and interact, we can observe the bridging feature connecting in the velocity space on the position-velocity diagram by comparing CO observations with synthetic observations \citep{2015MNRAS.450...10H,2015MNRAS.454.1634H,2017ApJ...835..142T,2017ApJ...850...23B}. We suggest that these results indicate the signature of the interaction between two molecular clouds in RCW 106 East and West.} 
{We comment on the viewing angle effects and expected velocity of the star-forming regions formed by colliding two clouds. 
\cite{2018ApJ...859..166F} argued the projection and viewing angle effects distort the contact interface layer on the position-velocity diagram comparing cloud-cloud collision in the Orion nebula cluster with the simulation model by \cite{2014ApJ...792...63T} and \cite{2015MNRAS.450...10H,2015MNRAS.454.1634H}.
In RCW106 East, the Higal compact sources have velocities around $-53$ \kms, and in RCW106 West, they have velocities around $-65$ \kms \citep{2021A&A...646A..74M}.
In the cloud-cloud collision model with internal turbulent motion, different sizes, and densities, the shock-compressed layer formed by momentum exchange does not necessarily have to be an intermediate velocity of the two colliding clouds (e.g., \citealp{2014ApJ...792...63T,2015MNRAS.454.1634H}). In addition, the velocities of massive clumps formed by cloud-cloud collision can have offsets from the intermediate velocity of two colliding clouds because shock compression enhances turbulence in the compressed layer \citep{2013ApJ...774L..31I,2021PASJ...73S.405F}. }

\begin{figure*}[h]
\begin{center} 
\includegraphics[width=18cm]{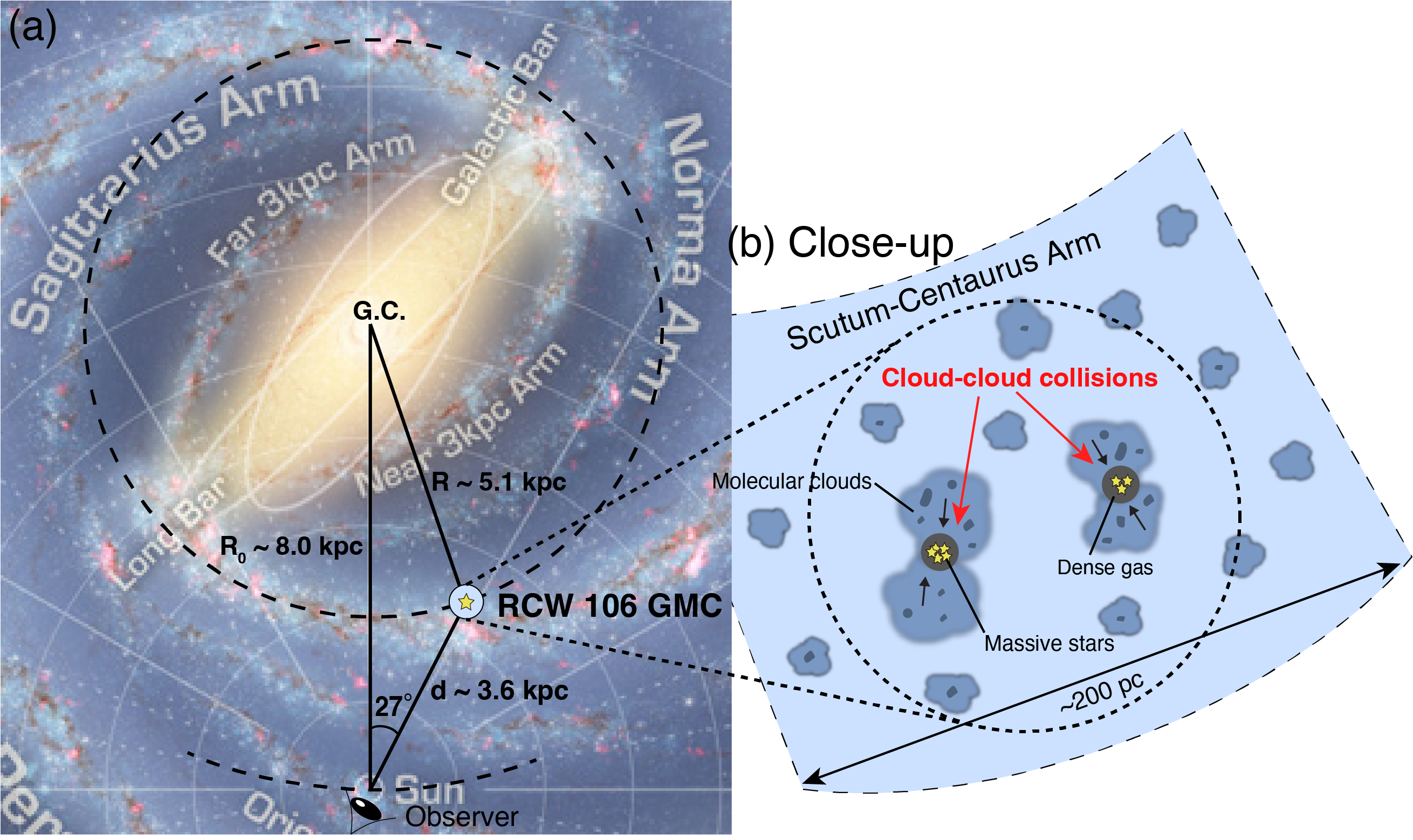}
\end{center}
\caption{(a) The location at the RCW 106 GMC complex superposed on the face-on illustration of the Milky Way (NASA/JPL-Caltech/ESO/R. Hurt). The distance to the Galactic center ($R_0$) is adopted from the Very Long Baseline Interferometry (VLBI) astrometry results \citep{2019ApJ...885..131R,2020PASJ...72...50V}. (b) Schematic illustration of our proposed massive star/cluster formation scenario in the RCW 106 GMC complex. }
\label{schematic}
\end{figure*}

Figure \ref{schematic} shows the schematic picture of our proposed scenario. 
Figure \ref{schematic}(a) illustrates the location of the RCW 106 GMC complex in the Scutum-Centaurus arm of the Milky Way. 
Molecular clouds, whose sizes and masses of $\sim$ 20--30 pc and $\sim 10^4$--$10^6 M_{\odot}$ collide with the supersonic velocity of $\sim 10$\kms, as shown in Figure \ref{schematic}(b).
The shock compression by supersonic collisions of molecular clouds produces dense cores and triggers massive star/cluster formation (see the review paper of \citealp{2021PASJ...73S...1F}). 
Indeed, the numerical simulations show that supersonic cloud-cloud collisions having different sizes and densities can produce massive molecular cloud cores required for massive star formation at the interface layer by shock compression \citep{1992PASJ...44..203H,2010MNRAS.405.1431A,2014ApJ...792...63T,2018PASJ...70S..58T,2018PASJ...70S..54S}.
Recent simulations considering magnetic fields also reveal massive dense cores produced by cloud-cloud collisions \citep{2013ApJ...774L..31I,2015ApJ...811...56W,2021PASJ...73S.405F,2021PASJ...73S.385S,2022ApJ...937...69K,2023ApJ...954..129K}. 
\cite{2018PASJ...70S..59K} point out that most cloud collision events occur between GMCs with the mass of $\lesssim 10^{5.5}\ M_{\odot}$ from their calculation for the time evolution of the GMC mass function.
The numerical simulation by \cite{2024MNRAS.52710077H} suggested that collisions of molecular clouds with masses of $\sim 10^4$--$10^5\ M_{\odot}$ at relative velocities of $\sim 10$--$20$\kms could occur in the spiral arm of the Milky Way-like galaxy.
These theoretical studies support our proposed scenario and observational signatures of the RCW 106 GMC complex.



\subsection{{Cloud-cloud collisions in the RCW 106 GMC complex}}
{In this section, we describe our cloud-cloud collision model in this region as the interpretation of observational results. The RCW 106 GMC complex has the orientation of star formation slightly inclined to the Galactic plane (see Figure 1a).
This might be explained by cloud collisions having the inclined momentum to the Galactic plane.
Our results of the morphology of the GMCs and their large projected distance could be explained by the Galactic scale dynamics such as spiral shocks or shear motions (e.g., \citealp{2014PASA...31...35D,2020ApJ...896...36T,2022PASJ...74...24K}).
Indeed, the Galactic-scale simulation of the cloud-cloud collision shows that the GMCs have elongated and filamentary structures due to Galactic dynamics (see Figure 14 in \citealp{2024MNRAS.52710077H}). These GMCs expect to experience the collision event in their lifetime.
RCW 106 East has more molecular mass than RCW 106 West, and the current evolved star formation indicate to be driven by the collision of more massive molecular clouds. RCW 106 West is still in the early stages of cloud collision and might develop into an extremely star formation in the future. 
In Figures \ref{ch}, \ref{east}, and \ref{west}, we can find that $-72$ km s$^{-1}$ MC$_{\rm EAST}$ and $-54$ km s$^{-1}$\ {MC$_{\rm WEST}$} appear spatially small distributions.
The observed these clumpy structures in RCW 106 East and West might be affected by two processes both molecular clouds are originally clumpy and cloud-cloud collisions form dense clumps.
The three-dimensional magnetohydrodynamic simulation shows that the colliding HI gas flow forms molecular clouds, and clumpy structures are created in the diffuse molecular gas \citep{2012ApJ...759...35I,2023ApJ...954...38K}.
The cloud-cloud collision scenario also forms dense clumps at the collision interface due to shock compression (e.g., \citealp{2014ApJ...792...63T,2018PASJ...70S..58T,2018PASJ...70S..53I}).}

\subsection{Comparison with properties of Galactic mini-starburst regions associated with the GMC complex.}
\begin{table*}[h]
\caption{Comparison with the Galactic mini-starbursts induced by cloud-cloud collisions in the GMC complex.}
\begin{tabular}{ccccccccccccc}
\hline
Name & $l$ & $b$ & Distance & Luminosity & Relative velocity & $N({\rm H_2})_{\rm max}$  & Molecular mass & References \\
&[degree] & [degree] & [kpc] & [$L_{\odot}$] & [\kms] & [cm$^{-2}$] & [$M_{\odot}$] &  \\
(1) & (2) & (3) &(4)& (5) & (6) & (7) & (8) & (9)  \\
\hline
RCW 106 East & $333.60$ & $-0.20$ & 3.6 & $4.8\times 10^6$ & $\sim 10$ & $\sim 3\times 10^{23}$ &$\sim 2\times 10^{6}$ &  This work\\ 
W33 & $12.80$ & $-0.20$ & 2.4 & $1.2\times 10^6$ & $\sim 23$ &$\sim 6\times 10^{23}$ & $\sim 9\times 10^5$& [1,2,3,4]\\
W43 Main & $30.80$ & $-0.05$ &  5.5 & $5.2\times 10^6$ & $\sim 10$-30  & $\sim 3\times 10^{23}$& $\sim 2\times 10^{6}$& [5,6,7]\\ 
W49A & $43.16$ & $0.00$ & 11.11 & $1.6\times 10^7$ & $\sim 8$  &$\sim 8\times 10^{23}$ &$\sim 1\times 10^6$& [8,9]\\
W51A & $49.5$ & $-0.04$ & 5.4 & $1.8\times 10^7$ & $\sim 4$--20 &$\sim 3\times 10^{23}$ &$\sim 6\times 10^5$& [10,11]\\
NGC 6334 & 351.20 & 0.70 & 1.7 & $2.7\times 10^6$ & $\sim 12$  & $\sim 8\times 10^{22}$& $\sim 2\times 10^5$ & [12,13]\\
NGC 6357 & 353.20 & 0.80 & 1.7 & $4.3\times 10^6$ & $\sim 12$  &$\sim 4\times 10^{22}$ & $\sim 2\times 10^5$ & [12,14]\\
Carina & 287.40 & -0.6 & 2.3 & $1.8\times 10^7$ &$\sim$6--20  & $\sim 5\times 10^{22}$& $\sim 8\times 10^{4}$ & [15,16,17]\\
\hline
\end{tabular}
  \vspace{3pt}
\label{phys_para}
{\raggedright Notes. --- Col. (1) Region name. Col. (2) Galactic longitude. Col. (3)  Galactic latitude. Col. (4) Distance from the solar system. Col. (5) Total infrared luminosity obtained by \cite{2018ApJ...864..136B}. Col. (6) The radial velocity difference between colliding molecular clouds. Col. (7) The peak H$_2$ column density. (8) Total molecular mass. (9) References. [1] \cite{2018PASJ...70S..50K} [2] \cite{2020MNRAS.496.1278D} [3] \cite{2022MNRAS.510.1106M} [4] \cite{2023MNRAS.519.2391Z} [5]\cite{2021PASJ...73S.129K} [6] \cite{2019PASJ...71S...1S} [7]\cite{2003ApJ...582..277M} [8] \cite{2022PASJ...74..128M} [9] \cite{2013ApJ...779..121G} [10] \cite{2021PASJ...73S.172F} [11] \cite{2001PASJ...53..793O} [12]\cite{2018PASJ...70S..41F} [13] \cite{2022A&A...660A..56A} [14] \cite{2019A&A...625A.134R} [15]\cite{2021PASJ...73S.201F} [16] \cite{2005ApJ...634..476Y} [17] \cite{2016MNRAS.456.2406R}
         \par}
\end{table*}
Finally, we compared the physical properties of molecular clouds associated with the Galactic mini-starbursts triggered by the cloud-cloud collision in the Milky Way.
Table \ref{phys_para} summarizes the properties of GMC complexes on the spiral arm in the Milky Way.
\cite{2021PASJ...73S..75E} conducted statistical studies on massive star-forming regions that reported cloud-cloud collisions. The authors suggest that the condition of $>10$ O-type star formation is the peak column density of $N(\rm {H_2})> 10^{23}$ cm$^{-2}$. Recent magnetohydrodynamic simulations of shock compression caused by colliding flows support this initial condition as a massive cluster formation \citep{2022ApJ...940..106A}.
We point out that RCW 106 East has a peak column density of $3 \times10^{23}$ cm$^{-2}$ and satisfies this condition.
{The relative velocity of $\sim 10$--20\kms are common to the mini-starburst regions produced by cloud-cloud collisions.}
Thus, we propose that the supersonic cloud collision events of $\sim 10$--20\kms are essential to achieve extreme star/cluster formation in the Milky Way.

The total molecular mass of RCW 106 East is $2\times 10^6\ M_{\odot}$.
It is massive compared with other GMCs on the spiral arm in Table \ref{phys_para}.
{Therefore, RCW 106 East is an agglomerate of molecular gas in the Sctutum-Centaurus arm and cloud-cloud collisions in this region are likely to be more frequent than the galactic disk's average.}
{The collision rate of spherical molecular clouds on the galactic disk is estimated as
\begin{equation}
t_{\rm coll} = {\lambda \over v_c} = {1 \over \pi r^2 n v_c},
\label{rate}
\end{equation}
where $\lambda$ is the mean free path, $r$ is the radius of molecular clouds, $n$ is the number density of molecular clouds, and $v_c$ is the relative velocity of a cloud-cloud collision, respectively. 
The number of clouds in the RCW 106 GMC complex is simply estimated to be $9.5\times10^6/7.0\times 10^5 \sim 13.6$ from the ratio of the $^{12}$CO total molecular mass in the GMC complex and mean molecular mass of the colliding clouds.
The number density is derived to be $n \sim 4.2 \times 10^{-6}\ {\rm pc}^{-3}$ from the number of clouds if we assume a volume space of 200 pc $\times$ 200 pc$ \times$ 80 pc.
The collision velocity is adopted as $v_c = 10 \times \sqrt{3} \sim 17$\kms assuming a viewing angle of 45\degr\ along the collision axis in the three-dimensional space. 
Taking a radius of molecular clouds of 20 pc, we obtain $t_{\rm coll}  \sim 11$ Myr. 
The total lifetime of GMCs is typically $\sim$10--30 Myr obtained by the statistical studies of GMCs in nearby galaxies \citep{2009ApJS..184....1K,2023ASPC..534....1C,2024arXiv240717018K,2024arXiv230519192}.
The collision rate of the RCW 106 GMC complex is $\sim 11$ Myr, which is comparable to the typical lifetime of GMCs. We point out that this result is likely to experience cloud collisional events more than once in its lifetime at this position in the Galaxy.
Indeed, numerical simulations also suggest that a massive GMC experiences several collisional events in its lifetime \citep{2009ApJ...700..358T,2011ApJ...730...11T,2014MNRAS.439..936F,2015MNRAS.446.3608D,2017ApJ...836..175K,2018PASJ...70S..59K}.
We suggest that cloud-cloud collisions predicted by these Galactic-scale simulations occur in RCW 106 GMC complex on the Scutum-Centaurus arm, and triggering extreme star formation events known as mini-starbursts.}

\section{Conclusion}
The summary of this paper is as follows:
\begin{enumerate}
\item We observed the southern massive star-forming region of RCW 106 (G333) GMC complex using a NANTEN2 4-m radio telescope in Chile operated by Nagoya University.
We also analyzed the Mopra CO Galactic plane survey and Herschel dust continuum data, which revealed the origin of the mini-starburst in the RCW 106 GMC complex.
\item {The RCW 106 GMC complex comprises two radial velocity components of $-68$\kms and $-50$\kms \citep{2015ApJ...812....7N}}. Using the Mopra CO data, our detailed analysis revealed three velocity components at the RCW 106 East and RCW 106 West, with the bright infrared dust emission involved in massive star/cluster formation.
\item {{The two out of three velocity components in RCW 106 East and RCW 106 West coincide CO peaks with infrared dust emission and connect each other with the bridge feature on the position-velocity diagram. Thus, these two velocity components} having a velocity difference of {$\sim$ 10\kms}\ are likely to be physically associated with massive star-forming regions and interact with each other in the RCW 106 GMC complex. We suggest that massive star/cluster formation in the RCW 106 GMC complex might be induced by supersonic cloud collisions in the Scutum-Centaurus arm. }
\item {The RCW 106 East has the peak H$_2$ column density of $N({\rm H_2}) > 10^{23}$ cm$^{-2}$. This column density is satisfied with the initial condition of $>10$ O-type star formation reported by previous studies of other massive star-forming regions \citep{2021PASJ...73S..75E}. The total mass of RCW 106 East is $\sim 2 \times 10^6\ M_{\odot}$, which is massive compared with other mini-starburst regions in the Milky Way. We point out that cloud-cloud collisions frequently occur in the RCW 106 GMC complex on the Scutum-Centaurus arm, which might trigger extreme star/cluster formation.}
\end{enumerate}
{We are grateful to the anonymous referee for carefully reading our manuscript and giving us thoughtful suggestions, which greatly improved this paper. }

{The NANTEN project is based on a mutual agreement between Nagoya University and 
the Carnegie Institution of Washington (CIW). We greatly appreciate the hospitality 
of all the staff members of the Las Campanas Observatory of CIW. We are thankful
to many Japanese public donors and companies who contributed to the realization 
of the project.}

NANTEN2 is an international collaboration of eleven universities: Nagoya University, Gifu University, Osaka Metropolitan University, University of Cologne, University of Bonn, Seoul National University, University of Chile, University of New South Wales, Macquarie University, University of Sydney, and Zurich Technical University.

The authors are grateful to Dr. Michael Burton of the University of New South Wales and the Armagh Observatory and Planetarium for the archival CO survey data with Mopra.
We also thank Dr. Graeme Wong for kindly supporting remote observations from Nagoya University.

The Mopra radio telescope is part of the Australia Telescope National Facility, which is funded by the Australian Government for operation as a National Facility managed by CSIRO. The University of New South Wales Digital Filter Bank used for the observations with the Mopra Telescope was provided with support from the Australian Research Council.

The Herschel spacecraft was designed, built, tested, and launched under a contract to ESA managed by the Herschel/Planck Project team by an industrial consortium under the overall responsibility of the prime contractor Thales Alenia Space (Cannes), and including Astrium (Friedrichshafen) responsible for the payload module and for system testing at spacecraft level, Thales Alenia Space (Turin) responsible for the service module, and Astrium (Toulouse) responsible for the telescope, with in excess of a hundred subcontractors.

PACS has been developed by a consortium of institutes led by MPE (Germany) and including UVIE (Austria); KU Leuven, CSL, IMEC (Belgium); CEA, LAM (France); MPIA (Germany); INAF-IFSI/OAA/OAP/OAT, LENS, SISSA (Italy); IAC (Spain). This development has been supported by the funding agencies BMVIT (Austria), ESA-PRODEX (Belgium), CEA/CNES (France), DLR (Germany), ASI/INAF (Italy), and CICYT/MCYT (Spain).

SPIRE has been developed by a consortium of institutes led by Cardiff University (UK) and including Univ. Lethbridge (Canada); NAOC (China); CEA, LAM (France); IFSI, Univ. Padua (Italy); IAC (Spain); Stockholm Observatory (Sweden); Imperial College London, RAL, UCL-MSSL, UKATC, Univ. Sussex (UK); and Caltech, JPL, NHSC, Univ. Colorado (USA). This development has been supported by national funding agencies: CSA (Canada); NAOC (China); CEA, CNES, CNRS (France); ASI (Italy); MCINN (Spain); SNSB (Sweden); STFC, UKSA (UK); and NASA (USA).

\vspace{5mm}
\facilities{NANTEN: 4 m, NANTEN2: 4 m, Mopra: 22 m, Herschel}
\software{Astropy \citep{2013A&A...558A..33A,2018AJ....156..123A,2022ApJ...935..167A}, NumPy \citep{2011CSE....13b..22V}, Matplotlib \citep{2007CSE.....9...90H}, IPython \citep{2007CSE.....9c..21P}, Miriad \citep{1995ASPC...77..433S}, and APLpy \citep{2012ascl.soft08017R}}

\appendix
\section{Procedures deriving the physical properties of molecular clouds}
We calculated the physical properties of molecular clouds from the $^{12}$CO and $^{13}$CO line intensity, assuming the CO-to-H$_2$ conversion factor ($X_{\rm CO}$) and Local Thermodynamic Equilibrium (LTE). 
We present their detailed methods in this appendix.

\subsection{The $X({\rm CO)}$ method}
The H$_2$ column density using $X({\rm CO)}$ is given by
\begin{equation}
N({\rm H_2})_{\rm 12x} = X_{\rm CO}\ I_{\rm ^{12}CO}\ {\rm [cm^{-2}]},
\label{XCO}
\end{equation}
where $I_{\rm ^{12}CO}$ is the integrated intensity of $^{12}$CO~$J$~=~1--0.
In this paper, we used $X_{\rm CO} = 2 \times 10^{20}\ {\rm [cm^{-2}\ (K\ km\ s^{-1})^{-1}]}$ as the standard value of the Galactic disk in the Milky Way \citep{2013ARA&A..51..207B}.

\subsection{The LTE method}
We derived the excitation temperature ($T_{\rm ex}$), optical depth ($\tau_{13}$), and column density of molecular clouds ($N_{\rm ^{13}CO}$) assuming the LTE as described in \cite{2008ApJ...679..481P,2020MNRAS.497.1851S}, and \cite{2024MNRAS.527.9290K,2024PASJ...76..579K}.
If we assume that the $^{12}$CO line is optically thick, the excitation temperature is given by
\begin{equation}
T_{\rm ex}=T_{0}^{115} \bigg/ {\rm ln} \left(1+\frac{T_{0}^{115}}{T_{\rm B}\ ({\rm ^{12}CO})_{\rm max}+{0.836}}\right)\ [{\rm K}],
\label{Tex}
\end{equation}
where $T_{\rm B} ({\rm ^{12}CO})_{\rm max}$ corresponds to the $^{12}$CO peak brightness temperature. $T_{0}^{115}= h\nu/k = {5.53}$ K is the Planck temperature with $h$, $\nu$, and $k$ being the Planck constant, rest frequency of $^{12}$CO~$J$~=~1--0, and Boltzmann constant, respectively. We assume that $T_{\rm ex}$ is equal in the $^{12}$CO and $^{13}$CO line emissions and express the optical depth as
\be
\tau_{13}=-{\rm ln}\left(1-\frac{T_{\rm B}({\rm ^{13}CO})_{\rm max}/T_{0}^{110}}{(e^{T_{0}^{110}/T_{\rm ex}}-1)^{-1}-{0.168}}\right),
\ee
where $T_{\rm B} ({\rm ^{13}CO})_{\rm max}$ and $T_{0}^{110}={5.29}$ K represent the $^{13}$CO peak {brightness temperature} and the Planck temperature at the rest frequency in $^{13}$CO~$J$~=~1--0, respectively. 
The $^{13}$CO column density is given by
\be
N_{\rm ^{13}CO}=3.0\times 10^{14} ~ \frac{\tau_{13}}{1-e^{-\tau_{13}}}
\frac{I_{\rm ^{13}CO}}{ 1-e^{-T_{0}^{110}/T_{\rm ex}}}\ [{\rm cm^{-2}}],
\ee
where $I_{\rm ^{13}CO}$ is the $^{13}$CO integrated intensity.
Then, we convert $N_{\rm ^{13}CO}$ to the H$_2$ column density using the abundance ratio of H$_2$ to $^{13}$CO molecules given by
\be
N({\rm H_2})_{\rm 13L}=Y_{\rm ^{13}CO} N_{\rm ^{13}CO}\ [{\rm cm^{-2}}].
\label{YN}
\ee
Here, $Y_{\rm ^{13}CO}$ is the abundance ratio adopted as $\sim 5\times 10^5$ \citep{1978ApJS...37..407D}.

Finally, we derived the total molecular mass of molecular clouds from the H$_2$ column densities obtained by the $X_{\rm CO}$ and LTE method.
The total molecular mass is given by
\begin{equation}
M = \mu_{\rm H_2} m_{\rm H} D^2 \sum_{i} \Omega\ N_i({\rm H_2})\ [M_{\odot}],
\label{mass}
\end{equation}
where $\mu_{\rm H_2}$, $m_{\rm H}$, $D$, $\Omega$, and $N_i(\rm{H_2})$ is the mean molecular weight per hydrogen molecule of 2.8, proton mass of $1.67 \times 10^{-24}$ g, distance from the solar system, solid angle, and the H$_2$ column density in each pixel, respectively.


\begin{thebibliography}{}
\bibitem[Abe et al.(2022)]{2022ApJ...940..106A} Abe, D., Inoue, T., Enokiya, R., et al.\ 2022, \apj, 940, 106. doi:10.3847/1538-4357/ac9e55

\bibitem[Abreu-Vicente et al.(2016)]{2016A&A...590A.131A} Abreu-Vicente, J., Ragan, S., Kainulainen, J., et al.\ 2016, \aap, 590, A131. doi:10.1051/0004-6361/201527674
\bibitem[Anathpindika(2010)]{2010MNRAS.405.1431A} Anathpindika, S.~V.\ 2010, \mnras, 405, 1431. doi:10.1111/j.1365-2966.2010.16541.x
\bibitem[Arzoumanian et al.(2022)]{2022A&A...660A..56A} Arzoumanian, D., Russeil, D., Zavagno, A., et al.\ 2022, \aap, 660, A56. doi:10.1051/0004-6361/202141699

\bibitem[Ascenso(2018)]{2018ASSL..424....1A} Ascenso, J.\ 2018, The Birth of Star Clusters, 424, 1. doi:10.1007/978-3-319-22801-3\_1

\bibitem[Astropy Collaboration et al.(2013)]{2013A&A...558A..33A} Astropy Collaboration, Robitaille, T.~P., Tollerud, E.~J., et al.\ 2013, \aap, 558, A33
\bibitem[Astropy Collaboration et al.(2018)]{2018AJ....156..123A} Astropy Collaboration, Price-Whelan, A.~M., Sip{\H{o}}cz, B.~M., et al.\ 2018, \aj, 156, 123. doi:10.3847/1538-3881/aabc4f
\bibitem[Astropy Collaboration et al.(2022)]{2022ApJ...935..167A} Astropy Collaboration, Price-Whelan, A.~M., Lim, P.~L., et al.\ 2022, \apj, 935, 167. doi:10.3847/1538-4357/ac7c74

\bibitem[Bains et al.(2006)]{2006MNRAS.367.1609B} Bains, I., Wong, T., Cunningham, M., et al.\ 2006, \mnras, 367, 1609. doi:10.1111/j.1365-2966.2006.10055.x
\bibitem[Becklin et al.(1973)]{1973ApJ...182L.125B} {Becklin, E.~E., Frogel, J.~A., Neugebauer, G., et al.\ 1973, \apjl, 182, L125. doi:10.1086/181234}

\bibitem[Binder \& Povich(2018)]{2018ApJ...864..136B} Binder, B.~A. \& Povich, M.~S.\ 2018, \apj, 864, 136. doi:10.3847/1538-4357/aad7b2
\bibitem[Bisbas et al.(2017)]{2017ApJ...850...23B} Bisbas, T.~G., Tanaka, K.~E.~I., Tan, J.~C., et al.\ 2017, \apj, 850, 23. doi:10.3847/1538-4357/aa94c5

\bibitem[Bonfand et al.(2024)]{2024A&A...687A.163B} {Bonfand, M., Csengeri, T., Bontemps, S., et al.\ 2024, \aap, 687, A163. doi:10.1051/0004-6361/202347856}

\bibitem[Bolatto et al.(2013)]{2013ARA&A..51..207B} Bolatto, A.~D., Wolfire, M., \& Leroy, A.~K.\ 2013, \araa, 51, 207. doi:10.1146/annurev-astro-082812-140944

\bibitem[Braiding et al.(2015)]{2015PASA...32...20B} Braiding, C., Burton, M.~G., Blackwell, R., et al.\ 2015, \pasa, 32, e020. doi:10.1017/pasa.2015.20

\bibitem[Braiding et al.(2018)]{2018PASA...35...29B} Braiding, C., Wong, G.~F., Maxted, N.~I., et al.\ 2018, \pasa, 35, e029. doi:10.1017/pasa.2018.18

\bibitem[Breen et al.(2007)]{2007MNRAS.377..491B} Breen, S.~L., Ellingsen, S.~P., Johnston-Hollitt, M., et al.\ 2007, \mnras, 377, 491. doi:10.1111/j.1365-2966.2007.11641.x


\bibitem[Burton et al.(2013)]{2013PASA...30...44B} Burton, M.~G., Braiding, C., Glueck, C., et al.\ 2013, \pasa, 30, e044. doi:10.1017/pasa.2013.22
\bibitem[Chevance et al.(2023)]{2023ASPC..534....1C} {Chevance, M., Krumholz, M.~R., McLeod, A.~F., et al.\ 2023, Protostars and Planets VII, 534, 1. doi:10.48550/arXiv.2203.09570}

\bibitem[Cubuk et al.(2023)]{2023PASA...40...47C} Cubuk, K.~O., Burton, M.~G., Braiding, C., et al.\ 2023, \pasa, 40, e047. doi:10.1017/pasa.2023.44

\bibitem[Dedes et al.(2011)]{2011A&A...526A..59D} Dedes, C., Leurini, S., Wyrowski, F., et al.\ 2011, \aap, 526, A59. doi:10.1051/0004-6361/200912874
\bibitem[Dell'Ova et al.(2024)]{2024A&A...687A.217D} {Dell'Ova, P., Motte, F., Gusdorf, A., et al.\ 2024, \aap, 687, A217. doi:10.1051/0004-6361/202348984}

\bibitem[Demachi et al.(2024)]{2024arXiv230519192} Demachi, F., Fukui, Y., Yamada, R.~I., et al.\ 2024, \pasj, 76, 1059. doi:10.1093/pasj/psae071

\bibitem[Dewangan et al.(2020)]{2020MNRAS.496.1278D} Dewangan, L.~K., Baug, T., \& Ojha, D.~K.\ 2020, \mnras, 496, 1278. doi:10.1093/mnras/staa1526

\bibitem[Dickman(1978)]{1978ApJS...37..407D} Dickman, R.~L.\ 1978, \apjs, 37, 407. doi:10.1086/190535

\bibitem[Dobbs \& Baba(2014)]{2014PASA...31...35D} {Dobbs, C. \& Baba, J.\ 2014, \pasa, 31, e035. doi:10.1017/pasa.2014.31}

\bibitem[Dobbs et al.(2015)]{2015MNRAS.446.3608D} Dobbs, C.~L., Pringle, J.~E., \& Duarte-Cabral, A.\ 2015, \mnras, 446, 3608. doi:10.1093/mnras/stu2319

\bibitem[Enokiya et al.(2021)]{2021PASJ...73S..75E} Enokiya, R., Torii, K., \& Fukui, Y.\ 2021, \pasj, 73, S75. doi:10.1093/pasj/psz119

\bibitem[Figuer{\^e}do et al.(2005)]{2005AJ....129.1523F} Figuer{\^e}do, E., Blum, R.~D., Damineli, A., et al.\ 2005, \aj, 129, 1523. doi:10.1086/427394

\bibitem[Fujimoto et al.(2014)]{2014MNRAS.439..936F} {Fujimoto, Y., Tasker, E.~J., Wakayama, M., et al.\ 2014, \mnras, 439, 936. doi:10.1093/mnras/stu014}

\bibitem[Fujita et al.(2021a)]{2021PASJ...73S.172F} Fujita, S., Torii, K., Kuno, N., et al.\ 2021a, \pasj, 73, S172. doi:10.1093/pasj/psz028
\bibitem[Fujita et al.(2021b)]{2021PASJ...73S.201F} Fujita, S., Sano, H., Enokiya, R., et al.\ 2021b, \pasj, 73, S201. doi:10.1093/pasj/psaa078

\bibitem[Fujiyoshi et al.(1998)]{1998MNRAS.296..225F} Fujiyoshi, T., Smith, C.~H., Moore, T.~J.~T., et al.\ 1998, \mnras, 296, 225. doi:10.1046/j.1365-8711.1998.01220.x
\bibitem[Fujiyoshi et al.(2001)]{2001MNRAS.327..233F} Fujiyoshi, T., Smith, C.~H., Wright, C.~M., et al.\ 2001, \mnras, 327, 233. doi:10.1046/j.1365-8711.2001.04704.x
\bibitem[Fujiyoshi et al.(2005)]{2005MNRAS.356..801F} Fujiyoshi, T., Smith, C.~H., Moore, T.~J.~T., et al.\ 2005, \mnras, 356, 801. doi:10.1111/j.1365-2966.2004.08501.x
\bibitem[Fujiyoshi et al.(2006)]{2006MNRAS.368.1843F} Fujiyoshi, T., Smith, C.~H., Caswell, J.~L., et al.\ 2006, \mnras, 368, 1843. doi:10.1111/j.1365-2966.2006.10255.x

\bibitem[Fukui et al.(2018a)]{2018ApJ...859..166F} {Fukui, Y., Torii, K., Hattori, Y., et al.\ 2018a, \apj, 859, 166. doi:10.3847/1538-4357/aac217}
\bibitem[Fukui et al.(2018b)]{2018PASJ...70S..41F} Fukui, Y., Kohno, M., Yokoyama, K., et al.\ 2018b, \pasj, 70, S41. doi:10.1093/pasj/psy017
\bibitem[Fukui et al.(2019)]{2019ApJ...886...14F} {Fukui, Y., Tokuda, K., Saigo, K., et al.\ 2019, \apj, 886, 14. doi:10.3847/1538-4357/ab4900}

\bibitem[Fukui et al.(2021a)]{2021PASJ...73S...1F} Fukui, Y., Habe, A., Inoue, T., et al.\ 2021a, \pasj, 73, S1. doi:10.1093/pasj/psaa103

\bibitem[Fukui et al.(2021b)]{2021PASJ...73S.405F} Fukui, Y., Inoue, T., Hayakawa, T., et al.\ 2021b, \pasj, 73, S405. doi:10.1093/pasj/psaa079

\bibitem[Galv{\'a}n-Madrid et al.(2013)]{2013ApJ...779..121G} Galv{\'a}n-Madrid, R., Liu, H.~B., Zhang, Z.-Y., et al.\ 2013, \apj, 779, 121. doi:10.1088/0004-637X/779/2/121
\bibitem[Gillespie et al.(1977)]{1977A&A....60..221G} {Gillespie, A.~R., Huggins, P.~J., Sollner, T.~C.~L.~G., et al.\ 1977, \aap, 60, 221}

\bibitem[Grave et al.(2014)]{2014A&A...563A.123G} Grave, J.~M.~C., Kumar, M.~S.~N., Ojha, D.~K., et al.\ 2014, \aap, 563, A123. doi:10.1051/0004-6361/201321306


\bibitem[Griffin et al.(2010)]{2010A&A...518L...3G} Griffin, M.~J., Abergel, A., Abreu, A., et al.\ 2010, \aap, 518, L3

\bibitem[Habe \& Ohta(1992)]{1992PASJ...44..203H} Habe, A. \& Ohta, K.\ 1992, \pasj, 44, 203

\bibitem[Haworth et al.(2015a)]{2015MNRAS.450...10H} Haworth, T.~J., Tasker, E.~J., Fukui, Y., et al.\ 2015a, \mnras, 450, 10. doi:10.1093/mnras/stv639
\bibitem[Haworth et al.(2015b)]{2015MNRAS.454.1634H} Haworth, T.~J., Shima, K., Tasker, E.~J., et al.\ 2015b, \mnras, 454, 1634. doi:10.1093/mnras/stv2068



\bibitem[Horie et al.(2024)]{2024MNRAS.52710077H} Horie, S., Okamoto, T., \& Habe, A.\ 2024, \mnras, 527, 10077. doi:10.1093/mnras/stad3798

\bibitem[Hunter(2007)]{2007CSE.....9...90H} Hunter, J.~D.\ 2007, Computing in Science and Engineering, 9, 90
\bibitem[Hyland et al.(1980)]{1980ApJ...241..709H} {Hyland, A.~R., McGregor, P.~J., Robinson, G., et al.\ 1980, \apj, 241, 709. doi:10.1086/158381}

\bibitem[Inoue \& Inutsuka(2012)]{2012ApJ...759...35I} Inoue, T. \& Inutsuka, S.-. ichiro .\ 2012, \apj, 759, 35. doi:10.1088/0004-637X/759/1/35

\bibitem[Inoue \& Fukui(2013)]{2013ApJ...774L..31I} Inoue, T. \& Fukui, Y.\ 2013, \apjl, 774, L31. doi:10.1088/2041-8205/774/2/L31

\bibitem[Inoue et al.(2018)]{2018PASJ...70S..53I} {Inoue, T., Hennebelle, P., Fukui, Y., et al.\ 2018, \pasj, 70, S53. doi:10.1093/pasj/psx089}
\bibitem[Karnik et al.(2001)]{2001MNRAS.326..293K} Karnik, A.~D., Ghosh, S.~K., Rengarajan, T.~N., et al.\ 2001, \mnras, 326, 293. doi:10.1046/j.1365-8711.2001.04596.x

\bibitem[Kashiwagi et al.(2023)]{2023ApJ...954..129K} Kashiwagi, R., Iwasaki, K., \& Tomisaka, K.\ 2023, \apj, 954, 129. doi:10.3847/1538-4357/ace7bd

\bibitem[Kawamura et al.(2009)]{2009ApJS..184....1K} Kawamura, A., Mizuno, Y., Minamidani, T., et al.\ 2009, \apjs, 184, 1. doi:10.1088/0067-0049/184/1/1
\bibitem[Kinoshita \& Nakamura(2022)]{2022ApJ...937...69K} Kinoshita, S.~W. \& Nakamura, F.\ 2022, \apj, 937, 69. doi:10.3847/1538-4357/ac8c95

\bibitem[Kobayashi et al.(2017)]{2017ApJ...836..175K} {Kobayashi, M.~I.~N., Inutsuka, S.-. ichiro ., Kobayashi, H., et al.\ 2017, \apj, 836, 175. doi:10.3847/1538-4357/836/2/175}

\bibitem[Kobayashi et al.(2018)]{2018PASJ...70S..59K} Kobayashi, M.~I.~N., Kobayashi, H., Inutsuka, S.-. ichiro ., et al.\ 2018, \pasj, 70, S59. doi:10.1093/pasj/psy018
\bibitem[Kobayashi et al.(2023)]{2023ApJ...954...38K} {Kobayashi, M.~I.~N., Iwasaki, K., Tomida, K., et al.\ 2023, \apj, 954, 38. doi:10.3847/1538-4357/ace34e}

\bibitem[Koda et al.(2016)]{2016ApJ...823...76K} Koda, J., Scoville, N., \& Heyer, M.\ 2016, \apj, 823, 76. doi:10.3847/0004-637X/823/2/76

\bibitem[Kohno et al.(2018)]{2018PASJ...70S..50K} Kohno, M., Torii, K., Tachihara, K., et al.\ 2018, \pasj, 70, S50. doi:10.1093/pasj/psx137
\bibitem[Kohno et al.(2021)]{2021PASJ...73S.129K} Kohno, M., Tachihara, K., Torii, K., et al.\ 2021, \pasj, 73, S129. doi:10.1093/pasj/psaa015
\bibitem[Kohno et al.(2022)]{2022PASJ...74...24K} {Kohno, M., Nishimura, A., Fujita, S., et al.\ 2022, \pasj, 74, 24. doi:10.1093/pasj/psab107}

\bibitem[Kohno \& Sofue(2024a)]{2024MNRAS.527.9290K} Kohno, M. \& Sofue, Y.\ 2024a, \mnras, 527, 9290. doi:10.1093/mnras/stad3648
\bibitem[Kohno \& Sofue(2024b)]{2024PASJ...76..579K} Kohno, M. \& Sofue, Y.\ 2024b, \pasj, 76, 579. doi:10.1093/pasj/psae033

\bibitem[Konishi et al.(2024)]{2024arXiv240717018K} Konishi, A., Muraoka, K., Tokuda, K., et al.\ 2024, \pasj, 76, 1098. doi:10.1093/pasj/psae073

\bibitem[Krause et al.(2020)]{2020SSRv..216...64K} {Krause, M.~G.~H., Offner, S.~S.~R., Charbonnel, C., et al.\ 2020, \ssr, 216, 64. doi:10.1007/s11214-020-00689-4}
\bibitem[Krumholz et al.(2019)]{2019ARA&A..57..227K} {Krumholz, M.~R., McKee, C.~F., \& Bland-Hawthorn, J.\ 2019, \araa, 57, 227. doi:10.1146/annurev-astro-091918-104430}

\bibitem[Kumar(2013)]{2013A&A...558A.119K} Kumar, M.~S.~N.\ 2013, \aap, 558, A119. doi:10.1051/0004-6361/201321627
\bibitem[Kutner \& Ulich(1981)]{1981ApJ...250..341K} Kutner, M.~L. \& Ulich, B.~L.\ 1981, \apj, 250, 341. doi:10.1086/159380
\bibitem[Lada \& Lada(2003)]{2003ARA&A..41...57L} Lada, C.~J. \& Lada, E.~A.\ 2003, \araa, 41, 57. doi:10.1146/annurev.astro.41.011802.094844
\bibitem[Ladd et al.(2005)]{2005PASA...22...62L} Ladd, N., Purcell, C., Wong, T., et al.\ 2005, \pasa, 22, 62. doi:10.1071/AS04068
\bibitem[LHAASO Collaboration(2024)]{2024arXiv240809905L} LHAASO Collaboration\ 2024, arXiv:2408.09905. doi:10.48550/arXiv.2408.09905
\bibitem[Lo et al.(2009)]{2009MNRAS.395.1021L} Lo, N., Cunningham, M.~R., Jones, P.~A., et al.\ 2009, \mnras, 395, 1021. doi:10.1111/j.1365-2966.2009.14594.x
\bibitem[Lo et al.(2011)]{2011MNRAS.415..525L} Lo, N., Redman, M.~P., Jones, P.~A., et al.\ 2011, \mnras, 415, 525. doi:10.1111/j.1365-2966.2011.18726.x
\bibitem[Lo et al.(2015)]{2015MNRAS.453.3245L} Lo, N., Wiles, B., Redman, M.~P., et al.\ 2015, \mnras, 453, 3245. doi:10.1093/mnras/stv1880

\bibitem[Lowe et al.(2014)]{2014MNRAS.441..256L} Lowe, V., Cunningham, M.~R., Urquhart, J.~S., et al.\ 2014, \mnras, 441, 256. doi:10.1093/mnras/stu568
\bibitem[Maity et al.(2024)]{2024arXiv240806826M} {Maity, A.~K., Inoue, T., Fukui, Y., et al.\ 2024, \apj, 974, 229. doi:10.3847/1538-4357/ad7098}
\bibitem[Maity et al.(2025)]{2024arXiv241113870M} {Maity, A.~K., Dewangan, L.~K., Bhadari, N.~K., et al.\ 2025, \aj, 169, 56. doi:10.3847/1538-3881/ad98ff}
\bibitem[M{\`e}ge et al.(2021)]{2021A&A...646A..74M} {M{\`e}ge, P., Russeil, D., Zavagno, A., et al.\ 2021, \aap, 646, A74. doi:10.1051/0004-6361/202038956}

\bibitem[Merello et al.(2013)]{2013ApJ...774...38M} Merello, M., Bronfman, L., Garay, G., et al.\ 2013, \apj, 774, 38. doi:10.1088/0004-637X/774/1/38
\bibitem[Miyawaki et al.(2022)]{2022PASJ...74..128M} Miyawaki, R., Hayashi, M., \& Hasegawa, T.\ 2022, \pasj, 74, 128. doi:10.1093/pasj/psab113
\bibitem[Mizuno \& Fukui(2004)]{2004ASPC..317...59M} Mizuno, A. \& Fukui, Y.\ 2004, Milky Way Surveys: The Structure and Evolution of our Galaxy, 317, 59
\bibitem[Mois{\'e}s et al.(2011)]{2011MNRAS.411..705M} Mois{\'e}s, A.~P., Damineli, A., Figuer{\^e}do, E., et al.\ 2011, \mnras, 411, 705. doi:10.1111/j.1365-2966.2010.17713.x

\bibitem[Molinari et al.(2010a)]{2010PASP..122..314M} Molinari, S., Swinyard, B., Bally, J., et al.\ 2010a, \pasp, 122, 314. doi:10.1086/651314
\bibitem[Molinari et al.(2010b)]{2010A&A...518L.100M} Molinari, S., Swinyard, B., Bally, J., et al.\ 2010b, \aap, 518, L100. doi:10.1051/0004-6361/201014659
\bibitem[Molinari et al.(2016)]{2016A&A...591A.149M} Molinari, S., Schisano, E., Elia, D., et al.\ 2016, \aap, 591, A149. doi:10.1051/0004-6361/201526380
\bibitem[Mookerjea et al.(2004)]{2004A&A...426..119M} Mookerjea, B., Kramer, C., Nielbock, M., et al.\ 2004, \aap, 426, 119. doi:10.1051/0004-6361:20040365

\bibitem[Motte et al.(2003)]{2003ApJ...582..277M} Motte, F., Schilke, P., \& Lis, D.~C.\ 2003, \apj, 582, 277. doi:10.1086/344538
\bibitem[Motte et al.(2018)]{2018ARA&A..56...41M} Motte, F., Bontemps, S., \& Louvet, F.\ 2018, \araa, 56, 41. doi:10.1146/annurev-astro-091916-055235
\bibitem[Motte et al.(2022)]{2022A&A...662A...8M} Motte, F., Bontemps, S., Csengeri, T., et al.\ 2022, \aap, 662, A8. doi:10.1051/0004-6361/202141677
\bibitem[Murase et al.(2022)]{2022MNRAS.510.1106M} Murase, T., Handa, T., Hirata, Y., et al.\ 2022, \mnras, 510, 1106. doi:10.1093/mnras/stab3472
\bibitem[Nguyen et al.(2015)]{2015ApJ...812....7N} Nguyen, H., Nguyen Lu'o'ng, Q., Martin, P.~G., et al.\ 2015, \apj, 812, 7. doi:10.1088/0004-637X/812/1/7
\bibitem[Nguyen Luong et al.(2011)]{2011A&A...529A..41N} Nguyen Luong, Q., Motte, F., Schuller, F., et al.\ 2011, \aap, 529, A41. doi:10.1051/0004-6361/201016271
\bibitem[Nguyen-Luong et al.(2016)]{2016ApJ...833...23N} Nguyen-Luong, Q., Nguyen, H.~V.~V., Motte, F., et al.\ 2016, \apj, 833, 23. doi:10.3847/0004-637X/833/1/23
\bibitem[Nishimura et al.(2020)]{2020SPIE11453E..3ZN} Nishimura, A., Ohama, A., Kimura, K., et al.\ 2020, \procspie, 11453, 114533Z. doi:10.1117/12.2562053
\bibitem[Okumura et al.(2001)]{2001PASJ...53..793O} Okumura, S.-I., Miyawaki, R., Sorai, K., et al.\ 2001, \pasj, 53, 793. doi:10.1093/pasj/53.5.793
\bibitem[Perez, \& Granger(2007)]{2007CSE.....9c..21P} Perez, F., \& Granger, B.~E.\ 2007, Computing in Science and Engineering, 9, 21
\bibitem[Pilbratt et al.(2010)]{2010A&A...518L...1P} Pilbratt, G.~L., Riedinger, J.~R., Passvogel, T., et al.\ 2010, \aap, 518, L1. doi:10.1051/0004-6361/201014759

\bibitem[Pineda et al.(2008)]{2008ApJ...679..481P} Pineda, J.~E., Caselli, P., \& Goodman, A.~A.\ 2008, \apj, 679, 481. doi:10.1086/586883

\bibitem[Pinheiro et al.(2012)]{2012MNRAS.423.2425P} Pinheiro, M.~C., Abraham, Z., Copetti, M.~V.~F., et al.\ 2012, \mnras, 423, 2425. doi:10.1111/j.1365-2966.2012.21049.x

\bibitem[Poglitsch et al.(2010)]{2010A&A...518L...2P} Poglitsch, A., Waelkens, C., Geis, N., et al.\ 2010, \aap, 518, L2. doi:10.1051/0004-6361/201014535

\bibitem[Pouteau et al.(2023)]{2023A&A...674A..76P} Pouteau, Y., Motte, F., Nony, T., et al.\ 2023, \aap, 674, A76. doi:10.1051/0004-6361/202244776

\bibitem[Rebolledo et al.(2016)]{2016MNRAS.456.2406R} Rebolledo, D., Burton, M., Green, A., et al.\ 2016, \mnras, 456, 2406. doi:10.1093/mnras/stv2776

\bibitem[Reid et al.(2009)]{2009ApJ...700..137R} Reid, M.~J., Menten, K.~M., Zheng, X.~W., et al.\ 2009, \apj, 700, 137. doi:10.1088/0004-637X/700/1/137
\bibitem[Reid et al.(2019)]{2019ApJ...885..131R} Reid, M.~J., Menten, K.~M., Brunthaler, A., et al.\ 2019, \apj, 885, 131. doi:10.3847/1538-4357/ab4a11

\bibitem[Ridge et al.(2006)]{2006AJ....131.2921R} Ridge, N.~A., Di Francesco, J., Kirk, H., et al.\ 2006, \aj, 131, 2921. doi:10.1086/503704
\bibitem[Robitaille \& Bressert(2012)]{2012ascl.soft08017R} Robitaille, T., \& Bressert, E.\ 2012, APLpy: Astronomical Plotting Library in Python, ascl:1208.017
\bibitem[Rodgers et al.(1960)]{1960MNRAS.121..103R} Rodgers, A.~W., Campbell, C.~T., \& Whiteoak, J.~B.\ 1960, \mnras, 121, 103. doi:10.1093/mnras/121.1.103
\bibitem[Romano et al.(2019)]{2019MNRAS.484.2089R} Romano, D., Burton, M.~G., Ashley, M.~C.~B., et al.\ 2019, \mnras, 484, 2089. doi:10.1093/mnras/sty3510
\bibitem[Russeil et al.(2005)]{2005A&A...429..497R} Russeil, D., Adami, C., Amram, P., et al.\ 2005, \aap, 429, 497. doi:10.1051/0004-6361:20048090
\bibitem[Russeil et al.(2016)]{2016A&A...587A.135R} {Russeil, D., Tig{\'e}, J., Adami, C., et al.\ 2016, \aap, 587, A135. doi:10.1051/0004-6361/201424484}

\bibitem[Russeil et al.(2019)]{2019A&A...625A.134R} Russeil, D., Figueira, M., Zavagno, A., et al.\ 2019, \aap, 625, A134. doi:10.1051/0004-6361/201833870

\bibitem[Sakre et al.(2021)]{2021PASJ...73S.385S} Sakre, N., Habe, A., Pettitt, A.~R., et al.\ 2021, \pasj, 73, S385. doi:10.1093/pasj/psaa059
\bibitem[Sasaki et al.(2024)]{2024A&A...682A.172S} Sasaki, M., Robrade, J., Krause, M.~G.~H., et al.\ 2024, \aap, 682, A172. doi:10.1051/0004-6361/202347154
\bibitem[Sault et al.(1995)]{1995ASPC...77..433S} {Sault, R.~J., Teuben, P.~J., \& Wright, M.~C.~H.\ 1995, Astronomical Data Analysis Software and Systems IV, 77, 433}
\bibitem[Shima et al.(2018)]{2018PASJ...70S..54S} Shima, K., Tasker, E.~J., Federrath, C., et al.\ 2018, \pasj, 70, S54. doi:10.1093/pasj/psx124
\bibitem[Simpson et al.(2012)]{2012MNRAS.419..211S} Simpson, J.~P., Cotera, A.~S., Burton, M.~G., et al.\ 2012, \mnras, 419, 211. doi:10.1111/j.1365-2966.2011.19686.x
\bibitem[Sofue et al.(2019)]{2019PASJ...71S...1S} Sofue, Y., Kohno, M., Torii, K., et al.\ 2019, \pasj, 71, S1. doi:10.1093/pasj/psy094
\bibitem[Sofue \& Kohno(2020)]{2020MNRAS.497.1851S} Sofue, Y. \& Kohno, M.\ 2020, \mnras, 497, 1851. doi:10.1093/mnras/staa2056
\bibitem[Takeuchi et al.(2010)]{2010PASJ...62..557T} Takeuchi, T., Yamamoto, H., Torii, K., et al.\ 2010, \pasj, 62, 557. doi:10.1093/pasj/62.3.557
\bibitem[Tamaoki et al.(2019)]{2019ApJ...875L..16T} Tamaoki, S., Sugitani, K., Nguyen-Luong, Q., et al.\ 2019, \apjl, 875, L16. doi:10.3847/2041-8213/ab1346
\bibitem[Takahira et al.(2014)]{2014ApJ...792...63T} Takahira, K., Tasker, E.~J., \& Habe, A.\ 2014, \apj, 792, 63. doi:10.1088/0004-637X/792/1/63
\bibitem[Takahira et al.(2018)]{2018PASJ...70S..58T} Takahira, K., Shima, K., Habe, A., et al.\ 2018, \pasj, 70, S58. doi:10.1093/pasj/psy011
\bibitem[Tasker(2011)]{2011ApJ...730...11T} Tasker, E.~J.\ 2011, \apj, 730, 11. doi:10.1088/0004-637X/730/1/11
\bibitem[Tasker \& Tan(2009)]{2009ApJ...700..358T} Tasker, E.~J. \& Tan, J.~C.\ 2009, \apj, 700, 358. doi:10.1088/0004-637X/700/1/358
\bibitem[Tokuda et al.(2019)]{2019ApJ...886...15T} {Tokuda, K., Fukui, Y., Harada, R., et al.\ 2019, \apj, 886, 15. doi:10.3847/1538-4357/ab48ff}
\bibitem[Tokuda et al.(2020)]{2020ApJ...896...36T} {Tokuda, K., Muraoka, K., Kondo, H., et al.\ 2020, \apj, 896, 36. doi:10.3847/1538-4357/ab8ad3}
\bibitem[Tokuda et al.(2022)]{2022ApJ...933...20T} {Tokuda, K., Minami, T., Fukui, Y., et al.\ 2022, \apj, 933, 20. doi:10.3847/1538-4357/ac6b3c}
\bibitem[Tokuda et al.(2023)]{2023ApJ...955...52T} {Tokuda, K., Harada, N., Tanaka, K.~E.~I., et al.\ 2023, \apj, 955, 52. doi:10.3847/1538-4357/acefb7}

\bibitem[Torii et al.(2017)]{2017ApJ...835..142T} Torii, K., Hattori, Y., Hasegawa, K., et al.\ 2017, \apj, 835, 142. doi:10.3847/1538-4357/835/2/142
\bibitem[Towner et al.(2024)]{2024ApJ...960...48T} {Towner, A.~P.~M., Ginsburg, A., Dell'Ova, P., et al.\ 2024, \apj, 960, 48. doi:10.3847/1538-4357/ad0786}

\bibitem[Ulich \& Haas(1976)]{1976ApJS...30..247U} Ulich, B.~L. \& Haas, R.~W.\ 1976, \apjs, 30, 247. doi:10.1086/190361

\bibitem[van der Walt et al.(2011)]{2011CSE....13b..22V} van der Walt, S., Colbert, S.~C., \& Varoquaux, G.\ 2011, Computing in Science and Engineering, 13, 22
\bibitem[Vall{\'e}e(2017)]{2017AstRv..13..113V} Vall{\'e}e, J.~P.\ 2017, The Astronomical Review, 13, 113. doi:10.1080/21672857.2017.1379459

\bibitem[VERA Collaboration et al.(2020)]{2020PASJ...72...50V} VERA Collaboration, Hirota, T., Nagayama, T., et al.\ 2020, \pasj, 72, 50. doi:10.1093/pasj/psaa018
\bibitem[Wakker \& van Woerden(1997)]{1997ARA&A..35..217W} Wakker, B.~P. \& van Woerden, H.\ 1997, \araa, 35, 217. doi:10.1146/annurev.astro.35.1.217
\bibitem[Wiles et al.(2016)]{2016MNRAS.458.3429W} Wiles, B., Lo, N., Redman, M.~P., et al.\ 2016, \mnras, 458, 3429. doi:10.1093/mnras/stw525
\bibitem[Wong et al.(2008)]{2008MNRAS.386.1069W} Wong, T., Ladd, E.~F., Brisbin, D., et al.\ 2008, \mnras, 386, 1069. doi:10.1111/j.1365-2966.2008.13107.x
\bibitem[Wu et al.(2015)]{2015ApJ...811...56W} Wu, B., Van Loo, S., Tan, J.~C., et al.\ 2015, \apj, 811, 56. doi:10.1088/0004-637X/811/1/56

\bibitem[Yang \& Wang(2020)]{2020A&A...640A..60Y} Yang, R.-Z. \& Wang, Y.\ 2020, \aap, 640, A60. doi:10.1051/0004-6361/202037518

\bibitem[Yonekura et al.(2005)]{2005ApJ...634..476Y} Yonekura, Y., Asayama, S., Kimura, K., et al.\ 2005, \apj, 634, 476. doi:10.1086/496869

\bibitem[Zhang et al.(2019)]{2019A&A...622A..52Z} Zhang, M., Kainulainen, J., Mattern, M., et al.\ 2019, \aap, 622, A52. doi:10.1051/0004-6361/201732400
\bibitem[Zhou et al.(2023)]{2023MNRAS.519.2391Z} Zhou, J.-W., Li, S., Liu, H.-L., et al.\ 2023, \mnras, 519, 2391. doi:10.1093/mnras/stac3559

\bibitem[Zhou et al.(2023)]{2023A&A...676A..69Z} Zhou, J.~W., Wyrowski, F., Neupane, S., et al.\ 2023, \aap, 676, A69. doi:10.1051/0004-6361/202346500
\bibitem[Zhou et al.(2024a)]{2024A&A...682A.128Z} Zhou, J.~W., Wyrowski, F., Neupane, S., et al.\ 2024a, \aap, 682, A128. doi:10.1051/0004-6361/202347377
\bibitem[Zhou et al.(2024b)]{2024arXiv240313442Z} Zhou, J.~W., Dib, S., Juvela, M., et al.\ 2024b, \aap, 686, A146. doi:10.1051/0004-6361/202449514

\bibitem[Zucker et al.(2018)]{2018ApJ...864..153Z} Zucker, C., Battersby, C., \& Goodman, A.\ 2018, \apj, 864, 153. doi:10.3847/1538-4357/aacc66
\end{thebibliography}

\end{CJK*}
\end{document}